\documentclass[useAMS,usenatbib]{mn2e}
\usepackage{graphicx}\usepackage{epsfig}\usepackage{epsf}
\usepackage{amsmath}\usepackage{amssymb}\usepackage{stfloats}
\usepackage{caption}

\title[Spectropolarimetry of SN 2011dh]{Spectropolarimetry of SN\,2011\lowercase{dh} in M51: geometric insights on a Type IIb supernova progenitor and explosion}
\author[Mauerhan et al.]{Jon C. Mauerhan$^{1}$\thanks{E-mail: mauerhan@astro.berkeley.edu}, G. Grant Williams$^{2,3}$, Douglas~C.~Leonard$^{4}$, Paul S. Smith$^{2}$, \newauthor Alexei V.\ Filippenko$^{1}$, Nathan Smith$^{2}$, Jennifer L. Hoffman$^{5}$, Leah Huk$^{5}$, \newauthor Kelsey~I.~Clubb$^{1}$, Jeffrey M. Silverman$^6$, S. Bradley Cenko$^{7}$, Peter Milne$^{2}$, \newauthor Avishay Gal-Yam$^{8}$, Sagi Ben-Ami$^{8}$\\ 
 $^{1}$Department of Astronomy, University of California, Berkeley, CA 94720-3411, USA \\
 $^{2}$Steward Observatory, University of Arizona, 933 N. Cherry Ave., Tucson, AZ 85721, USA \\
 $^{3}$MMT Observatory, Tucson, AZ 85721-0065, USA \\
 $^{4}$Department of Astronomy, San Diego State University, PA-210, 5500 Campanile Drive, San Diego, CA 92182-1221 \\
$^{5}$Department of Physics \& Astronomy, University of Denver, 2112 East Wesley Avenue, Denver, CO 80208 \\
 $^{6}$Department of Astronomy, University of Texas at Austin, Austin, TX 78712, USA \\
$^{7}$NASA Goddard Space Flight Center, Greenbelt, MD 20771, USA \\
$^{8}$Department of Particle Physics and Astrophysics, Weizmann Institute of Science, Rehovot 76100, Israel } 
\begin{document}

\pagerange{\pageref{firstpage}--\pageref{lastpage}} \pubyear{2013}
\maketitle
\label{firstpage}

\begin{abstract}
We present seven epochs of spectropolarimetry of the Type IIb supernova (SN) 2011dh in M51, spanning 86 days of its evolution. The first epoch was obtained 9 days after the explosion, when the photosphere was still in the depleted hydrogen layer of the stripped-envelope progenitor. Continuum polarization is securely detected at the level of $P\approx0.5$\% through day 14 and appears to diminish by day 30, which is different from the prevailing trends suggested by studies of other core-collapse SNe. Time-variable modulations in $P$ and position angle are detected across P-Cygni line features. H$\alpha$ and He\,{\sc i} polarization peak after 30 days and exhibit position angles roughly aligned with the earlier continuum, while O\,{\sc i} and Ca\,{\sc ii} appear to be geometrically distinct. We discuss several possibilities to explain the evolution of the continuum and line polarization, including the potential effects of a tidally deformed progenitor star, aspherical radioactive heating by fast-rising plumes of $^{56}$Ni from the core, oblique shock breakout, or scattering by circumstellar material. While these possibilities are plausible and guided by theoretical expectations, they are not unique solutions to the data. The construction of more detailed hydrodynamic and radiative-transfer models that incorporate complex aspherical geometries will be required to further elucidate the nature of the polarized radiation from SN\,2011dh and other Type IIb supernovae. 
\end{abstract}
\begin{keywords}
supernovae: general --- supernovae: individual (SN~2011dh)
\end{keywords}

\section{Introduction}
Spectropolarimetric observations of supernovae (SNe) probe the explosion geometry and the relative distribution of chemical elements within the progenitor's outer envelope and inner ejecta (McCall 1984; see also the review by Wang \& Wheeler 2008). This is valuable information that can provide clues to the nature of the progenitor systems, the explosion mechanism, and the kicks imparted to newborn neutron stars.  

The most common intrinsic source of linearly polarized emission in SNe is Thomson scattering of photons by free electrons in the dense ionized outflow or circumstellar medium. Optical scattering of photons by circumstellar dust particles is another potentially important mechanism for efficiently producing linear polarization (Wang \& Wheeler 1996). However, in the idealised case of spherically symmetric geometry, every electric-field vector from the circular scattering surface, as observed on the sky, will have an orthogonally oriented vector of equal magnitude in an adjacent quadrant of the circle, and will thus cancel out. The detection of net polarization therefore requires some degree of asphericity on the plane of the sky to break the symmetry. Net {\it continuum} polarization indicates the presence of global asphericity for the electron-scattering photosphere and/or circumstellar medium, or a patchy distribution of continuum opacity.  Polarization changes across {\it line} features, on the other hand, can occur for both globally spherical and aspherical configurations, when there is partial obscuration of the underlying photosphere by an uneven or clumpy distribution of line opacity (Kasen et al. 2003; Dessart \& Hillier 2011). These general features are potentially important for both thermonuclear and core-collapse (CC) explosions. However, in order for spectropolarimetry to provide information on the intrinsic asphericity of SNe, one must have knowledge of the wavelength-dependent linear polarization produced by the differential absorption of light by interstellar dust along the line of sight (Serkowski 1968), from both the Milky Way and the host galaxy.

A large fraction of SNe that have been observed with polarimetry exhibit evidence for some degree of asphericity, but the spectropolarimetric properties are wide ranging. As studies have progressed in the last decade, systematic trends have emerged for the various SN subtypes. Thermonuclear SNe\,Ia generally exhibit a relatively low degree of intrinsic net polarization in the continuum ($P \lesssim 0.2$\%; Wang \& Wheeler 2008; Wang et al. 2003a; Kasen et al. 2003) and more  substantial polarization across P-Cygni absorption features (occasionally  $\sim1$\% or more; e.g., Leonard et al. 2005; Patat et al. 2009), indicating only modest asphericity for the overall geometry of the outflow, yet a sometimes highly nonuniform distribution of line optical depth.  CC~SNe have exhibited markedly different evolution. Some of the most well-studied SNe\,II-P have exhibited an evolving degree of continuum polarization, beginning at low or modest levels during the plateau phase and then sharply increasing as the photosphere recedes below the H envelope and into the more highly aspherical He layer and metal-rich ejecta (Leonard et al. 2006; Chornock et al. 2010; but also see Leonard et al. 2012, and preliminary results of Leonard et al. 2013). SNe\,IIn, which are explosions that interact with circumstellar material (CSM) to produce their namesake relatively {\it narrow} emission lines, commonly show a moderately high degree of continuum polarization at early times (1--2\%; Leonard et al. 2000; Hoffman et al. 2008; Smith et al. 2008; Patat et al. 2011; Mauerhan et al. 2014), and in some cases exhibit a complex temporal evolution in polarization magnitude and position angle on the sky (Mauerhan et al. 2014). These characteristics have been interpreted as evidence for a superposition of multiple, geometrically distinct components of polarization from the SN outflow and a highly aspherical distribution of shocked CSM, consistent with toroidal geometry in several cases (Hoffman et al. 2008; Mauerhan et al. 2014). 

Stripped-envelope SNe are explosions that stem from stars that have depleted hydrogen envelopes (Type IIb) or that are entirely devoid of hydrogen (Type Ib) and helium (Type Ic) lines; see Filippenko (1997) for a review of SN classification. These subclasses of SNe consistently exhibit substantial polarization in the continuum and enhanced polarization across lines (Leonard et al. 2002; Wang et al. 2003b; Wang \& Wheeler 2008). SNe\,IIb, in particular, are an intermediate class between SNe\,II-P and Ib, and constitute $\sim11$\% of CC~SNe in large Milky Way-like galaxies (Smith et al. 2011; Li et al. 2011). In the majority of cases, the stellar progenitors at the time of explosion are thought to have been yellow supergiants or Wolf-Rayet stars of relatively modest initial mass (10--20\,M$_\odot$) that had the majority of their hydrogen envelopes stripped down to $M_{\textrm \tiny{H}} \approx 0.001$--1\,M$_{\odot}$ via gravitational interaction with a binary companion (e.g., Podsiadlowski et al. 1993; Woosley et al. 1994; Utrobin 1994; H{\"o}flich 1995; Maund et al. 2004; Claeys et al. 2011; Smith et al. 2011; Van Dyk et al. 2013; Gal-Yam et al. 2014; Jerkstrand et al. 2015), in some cases potentially forming stripped-envelope Wolf-Rayet stars before exploding. These hypotheses are supported by the direct detection of the progenitor stars and their candidate companions in respective pre- and post-explosion images from the \textit{Hubble Space Telescope (HST)} (Maund et al. 2004; Van Dyk et al. 2011, 2013; Fox et al. 2014). 

Chevalier \& Soderberg (2010) suggested that SNe\,IIb be subcategorized into two types of explosions that stem from either (i) progenitors with extended atmospheres and slow dense winds (eIIb; $R \approx 100$\,R$_\odot$), leading to a prominent early peak in the light curve from the extended shock-heated envelope, or (ii) more compact progenitors with fast tenuous winds (cIIb; $R \approx 1$\,R$_\odot$). The latter subclass appears to be distinguishable by weak optical emission from the shock-heated envelope at early times, nonexistent or very weak H emission in the late nebular phase, rapid radio evolution, and nonthermal X-ray emission. The cIIb class is thus more consistent with Wolf-Rayet-like progenitor stars than cool supergiants having extended atmospheres. The classification of SNe\,IIb within these subcategories is not always clear, however, as there could be a continuum of intermediate objects having progenitor radii between the eIIb and cIIb extremes (Horesh et al. 2013). Different amounts of ultraviolet (UV) excess might also be an important diagnostic for determining the nature of SN\,IIb progenitors and the properties of their CSM (Ben-Ami et al. 2015).

At early phases, SNe\,IIb display Balmer-dominated spectra and commonly exhibit a low to moderate-level continuum polarization (see the review by Wang \& Wheeler et al. 2008, and references therein), indicative of modestly aspherical global outflow. By the time of maximum light, just weeks after explosion, the photospheres of SNe\,IIb have already passed through the depleted hydrogen layer, giving rise to He-dominated spectra; this IIb$\rightarrow$Ib transition is often accompanied by a substantial increase in \textit{line} polarization. Well-studied examples for which quality spectropolarimetry has been obtained include the SN\,IIb prototype SN\,1993J (Trammel et al. 1993; H{\"o}flich 1995; H{\"o}flich et al. 1996; Tran et al. 1997) and the remarkably similar SN\,1996cb (Wang et al. 2001), both of which exhibited substantial continuum polarizations of $\sim0.6$\% before peak brightness and H-rich phases that lasted for 2--3 weeks after the explosion. The spectropolarimetric similarity between these SNe is noteworthy, given that SN\,1993J has been classified as an eIIb and SN\,1996cb as a cIIb (Chevalier \& Soderberg 2010). The additional case of the Type IIb SN\,2001ig began with a low ($\sim0.2$\%) continuum polarization at early times, and later rose to $\sim1$\% after the photosphere receded into the He layer (Maund et al. 2007a), similar in behaviour to SNe\,II-P. The more recent cIIb SN\,2008ax was a rather unusual object that exhibited the Balmer-dominated spectrum of a young SN\,IIb only during the first several days after exploding, indicative of an H envelope mass that was substantially lower than that of most other SNe\,IIb (e.g., see Marion et al. 2014). The object exhibited substantial continuum polarization of $\sim0.6$\% at early times, indicative of an aspherical outer envelope, and developed very strong H$\alpha$ line polarization of $\sim3.4$\%, stronger than ever observed in any SN\,IIb and with large radial velocities reaching 20,000\,km\,s$^{-1}$ (Chornock et al. 2011).  

Overall, spectropolarimetric properties of SNe\,IIb are broadly consistent with a moderately aspherical photosphere, punctuated by irregular clumps or plumes of enhanced line optical depth in the ejecta. Optical scattering of light from CSM dust particles is also a potentially important factor affecting the degree of continuum polarization, and was possibly influential in the case of the prototypical SN\,1993J (Tran et al. 1997).

SN\,2011dh was a bright Type IIb explosion in the nearby spiral galaxy NGC\,5194 (Messier 51a, hereafter referred to as M51), discovered independently by several amateur astronomers (summarized by Griga et al. 2011 and Arcavi et al. 2011). Remarkably, a yellow supergiant (YSG) star was found to be spatially coincident with the SN in {\it HST} archival images of the galaxy (Li \& Filippenko 2011; Maund et al. 2011; Van Dyk et al. 2011). It was unclear whether this star was the actual progenitor of the explosion, a neighbor, or even a surviving companion star to a more compact progenitor as initially hypothesized on the basis of its early-time luminosity (Arcavi et al. 2011) and radio emission (Soderberg et al. 2012). Eventually, the identification of the YSG as the true progenitor was solidified after the SN had faded, revealing that the YSG star had disappeared (Van Dyk et al. 2013). The same star was found to be variable in ground-based images up to 3\,yr before the explosion (Szczygie{\l} et al. 2012). Recently, a candidate hot companion star has been detected in late-time UV data from \textit{ HST} (Folatelli et al. 2014); SN\,2011dh is thus the second SN\,IIb to exhibit direct evidence for a binary companion, after SN\,1993J (Maund et al. 2004; Fox et al. 2014). 

Estimates of the YSG progenitor's initial main-sequence mass and radius lie in the range 10--19\,M$_{\odot}$ and  200--300\,R$_{\odot}$, respectively (Sahu, Anupama, \& Chakradhari 2013; Van Dyk et al. 2011; Bersten et al. 2012; Benvenuto et al. 2013; Ergon et al. 2014); SN\,2011dh is thus a member of the eIIb subclass. The eIIb designation is also consistent with the photometric detection of the shock-heated envelope phase in the early-time light curve (Arcavi et al. 2011), although the fact that this phase was relatively short lived compared to that of SN\,1993J suggests that the progenitor of SN\,2011dh was substantially less extended. Spectroscopic evolution of SN\,2011dh showed an early H-rich phase lasting for roughly two weeks, after which the photosphere entered the He layer (Arcavi et al. 2011; Marion et al. 2014; Ergon et al. 2014). The late-time decline in the light curve showed the explosion to be consistent with the production of 0.09\,M$_{\odot}$ of $^{56}$Ni and 0.2\,M$_{\odot}$ of oxygen, and also provided evidence for possible late-time dust formation, CSM interaction, and substantial asphericities in the explosion ejecta (Sahu et al. 2013; Shivvers et al. 2013). Moreover, recent high-resolution VLBI imaging at 8.4\,GHz might have detected ``hot spots" potentially indicative of modestly aspherical distribution of the outer ejecta, although this feature might just be an artifact of the data-reduction process (de Witt et al. 2015). Spectropolarimetry of SN\,2011dh is thus well motivated. 

Here we present high-quality multi-epoch spectropolarimetry of SN\,2011dh, producing the most complete dataset of its kind for a SN\,IIb. In \S2 the observational facilities are described. We present the data and analyze the results in \S3.  Section 4 compares SN\,2011dh to other SNe\,IIb that have been the subject of spectropolarimetric observations and interprets the data in the context of existing explosion and progenitor models.

\section{OBSERVATIONS}

\subsection{The Kast Spectrograph at Lick Observatory}

The Shane 3\,m reflector at Lick Observatory equipped with the Kast spectrograph (Miller \& Stone 1993) was used to observe SN\,2011dh on 2011 June 6 and 29, and on August 4 and 25 (UT dates are used throughout this paper). Kast is a dual-beam spectropolarimeter that utilizes a rotatable semiachromatic half-waveplate to modulate the incident polarization and a Wollaston prism in the collimated beam to separate the two orthogonally polarized spectra onto the CCD detector.  Only the red channel of Kast was used for spectropolarimetry; a GG455 order-blocking filter suppressed all second-order light at wavelengths shortward of 9000\,{\AA}. Observations were made with the 300\,line\,mm$^{-1}$ grating and the 3{\arcsec}-wide slit, yielding a spectral resolution of $\sim4.6$\,{\AA}\,pixel$^{-1}$ and a full width at half-maximum intensity (FWHM) spectral resolution of $\sim16$\,{\AA} (as determined by the FWHM of lamp lines). The useful wavelength coverage with this setup is 4600--9000\,{\AA}. The orientation of the slit on the sky was always set to a position angle of $180^{\circ}$ (i.e., aligned north-south). Exposures of 900\,s were obtained at each of four waveplate positions ($0\fdg0$, $45\fdg0$, $22\fdg5$, and $67\fdg5$). In most cases, three waveplate sequences were performed and the results median combined. Flatfield and arc lamp (for wavelength calibration) spectra were obtained immediately after each sequence, without moving the telescope. The arc lamps are internal to the Kast instrument, while the flatfield spectra were produced by the reflection of incandescent lamp light off of the inner surface of the dome, which was rotated into the field of view of the telescope before or after the science exposures. The flatfield spectral images were normalized by a low-order spline fit to the continuum before dividing them into the science images.

For polarimetric calibrations, standard stars were selected from the sample of Schmidt et al. (1992a,b). The unpolarized standard stars BD+26$^{\circ}$2606 and BD+32$^{\circ}$3739 were observed to verify the low instrumental polarization of the Kast spectrograph. We constrained the average fractional Stokes $Q$ and $U$ values to $<0.05$\%. By observing the above unpolarized standard stars through a 100\% polarizing filter, we determined that the polarimetric response is so close to 100\% that no correction was necessary, and we obtained the instrumental polarization position-angle curve, which we used to correct the data. We observed the high-polarization stars HD\,154445, HD\,127769, HD\,161056, and HD\,204827 to obtain the zeropoint of the polarization position angle on the sky ($\theta$) and to determine the accuracy of polarimetric measurements, which were generally consistent with previously published values to within $\Delta P<0.05$\% and $\Delta \theta<1^{\circ}$. 

 \begin{center}
 \begin{table*} 
      \caption{Integrated $V$-band and continuum polarization of SN~2011dh}
\renewcommand\tabcolsep{5.5pt} \scriptsize
\begin{tabular}[b]{@{}llrccrrr}
\hline
\hline
UT Date & MJD &day$^{\textrm{a}}$ &Tel./Instr.   &  $P_{V}$\,(\%)$^{\textrm{b}}$ & $\theta_{V}$\,(deg)$^{\textrm{b}}$ &  $P_{\rm cont}$\,(\%)$^{\textrm{e}}$ & $\theta_{\rm cont}$\,(deg)$^{\textrm{f}}$ \\
 \hline
  \hline
Jun. 09 & 5721.8              & 9               &Lick/Kast               & 0.406 (0.010)  & 34.9 (0.8)                            & 0.47 (0.02) &  23.5 (0.8) \\
Jun. 14$^\textrm{c}$ & 5727          & 14              &Bok/SPOL         &  0.237 (0.013) &  14.8 (1.0)          & 0.45 (0.02) &  2.3 (1.3)  \\
Jun. 28 & 5742.8             &  30                &Lick/Kast             & 0.358 (0.013)  & 8.7 (1.0)                             & 0.18 (0.04) &  12.0 (3.4) \\
Jul. 06 &  5748.7             &  36               &P200/DBSpec      & 0.345 (0.021)  &  18.7 (1.7)                         & 0.21 (0.03) &  29.1 (2.7) \\
Jul. 28$^\textrm{d}$ &  5771             &  58                &Bok/SPOL             & 0.425 (0.042)  & 18.3 (4.2) & -~~~~~~~~& -~~~~~~~~\\ 
Aug. 04 & 5777.7               &  65               &Lick/Kast              &  0.400 (0.022) & 13.2 (1.7)                         & -~~~~~~~ &-~~~~~~~~\\
Aug. 25 & 5798.7           &  86                 &Lick/Kast              &  0.259 (0.035) & 15.2 (3.9)                            &    -~~~~~~~           &  -~~~~~~~~\\
\hline
\end{tabular}\label{tab:p48} 
\begin{flushleft}
 \scriptsize$^\textrm{a}$Day is with respect to the adopted explosion date (JD~2455713.08), and rounded to the nearest integer value. \\
 \scriptsize$^\textrm{b}$$V$-band values averaged over the wavelength range 5050--5950\,{\AA}. \\
 \scriptsize$^\textrm{c}$Average of data from June 14 and June 15. \\
 \scriptsize$^\textrm{d}$Average of data from  July 28 and July 30. \\
 \scriptsize$^\textrm{e}$Average of data from  Aug 4 and Aug 5. \\
 \scriptsize$^\textrm{f}$Continuum sample region is 7000--7450\,{\AA}; some weak line contamination might be present for the June 28 and July 6 data. \\
\end{flushleft}
\end{table*}
\end{center}

\subsection{SPOL at Arizona Observatories}
Spectropolarimetric data were obtained on 2011 June 14, 15, 28, and 30 with the Bok 90\,inch telescope on Kitt Peak. Observations made use of the CCD Imaging/Spectropolarimeter (SPOL;  Schmidt et al. 1992a). Like Kast, SPOL contains a rotatable semiachromatic half-waveplate that is used to modulate incident polarization and a Wollaston prism in the collimated beam to separate the two orthogonally polarized spectra onto the detector. SPOL was configured with a 600\,line\,mm$^{-1}$ grating in first order.  The slit selection was based on the seeing conditions and ranged from $1.1\arcsec$ to $5.1\arcsec \times 51\arcsec$.  This setup yielded a spectral resolution of 20--30\,{\AA} (FWHM) with useful wavelength coverage in the range 4000--7600\,{\AA}.  A standard Hoya L38 blocking filter was used to ensure that the first-order spectrum was not contaminated by second-order light for $\lambda \gtrsim 7600$\,\AA. The efficiency of the waveplate as a function of wavelength was measured and corrected at all epochs by inserting a fully polarizing Nicol prism into the beam above the slit. The orientation of the slit on the sky was always set to a position angle of $0^{\circ}$ (i.e., aligned north-south). A series of four separate exposures that sample 16 orientations of the waveplate yields two independent, background-subtracted measures of each of the normalized linear Stokes parameters, $q$ and $u$. For some observing runs, several polarization sequences of SN~2011dh from different nights were combined if no polarimetric change was discernible, with the weighting of the individual measurements based on photon statistics. 

The instrumental polarization of Bok/SPOL is very small ($<0.1$\%), as measurements of unpolarized standard stars over the past two decades have consistently shown. Still, to confirm, we observed one or both of the unpolarized standard stars BD+28$^{\circ}$4211 and G191B2B (Schmidt et al. 1992b) during each epoch, confirming that the instrumental polarization is $<0.1$\% and that there is no low-level residual structure observed in the $q$ and $u$ Stokes spectra.  The linear polarization position angle on the sky ($\theta\/$) was determined by observing one or more of the interstellar polarization standards Hiltner~960, VI~Cyg~\#12, or BD+59$^{\circ}$389 (Schmidt et al. 1992b) during all epochs. The adopted correction from the instrumental to the standard equatorial frame for $\theta\/$ for all epochs was determined from the average position angle offset of the polarized standards. Differences between the measured and expected polarization position angles were $<0\fdg3$ for all of the standard stars.

\subsection{The Double Spectrograph at Palomar Observatory}

With the 5\,m Hale telescope at Palomar Observatory, we obtained a
single epoch of spectropolarimetry of SN\,2011dh on 2011 July 6, using
the Double Spectrograph (DBSP; Oke \& Gunn 1982) with polarimeter.
The total exposure time was 5760\,s, spanning four complete sets of
spectropolarimetric data (i.e., each set consisted of four 360\,s
observations at each of four waveplate position angles). 
 
Owing to spurious polarization features being introduced by the
dichroic (see, e.g., Ogle et al. 1999), we observed with only the red
camera of the spectrograph (i.e., no dichroic) using the 158/7500
grating and the GG445 order-blocking filter (to mitigate second-order
light contamination).  This resulted in a final, useful wavelength
range of 5100--9100\,{\AA}, with a spectral resolution of
$\sim15$\,{\AA} (FWHM). We note that the absence of a dichroic
resulted in poor spatial and spectral focus in one of the beams (the
``top'' beam of the two beams delivered by the spectropolarimeter on
the CCD chip) blueward of 7500\,{\AA}, with the resolution approaching
$\sim30$\,{\AA} near the blue end of the spectral range. In order to
preserve the best spectral resolution, the total-flux spectrum was
produced using only the bottom beam in the wavelength range
5100--7500\,{\AA}, with the two beams being combined redward of
7500\,{\AA}.  For the spectropolarimetry, we retain the entire
spectral range, and so the resolution of any spectropolarimetric
features varies somewhat from the blue end to the red end of the
spectrum.

To derive the total-flux spectra, we extracted all one-dimensional
sky-subtracted spectra optimally (Horne 1986) in the usual manner.  Each
spectrum was then wavelength and flux calibrated, and was corrected for
continuum atmospheric extinction and telluric absorption bands (e.g., Matheson
et al. 2000).  The total-flux spectrum was calibrated by observing the flux
standard star BD+17$^\circ4708$; note, however, that neither SN~2011dh nor the
flux standard were observed at the parallactic angle (Filippenko 1982), and so
the relative-flux calibration of the spectrum is somewhat suspect. 

The polarization-angle offset between the half-wave plate and the sky
coordinate system was determined by observing the polarized standard star
HD\,204827 and setting the $R$-band polarization position angle (i.e.,
$\theta_R$, the debiased, flux-weighted average of the polarization angle over
the wavelength range 6100--8100\,{\AA}; see Leonard et al. 2001) to be
$59.1^\circ$, the value cataloged by Schmidt et al. (1992a).  With the
zeropoint now defined, we also measured the polarization position angle of the
polarized standard star HD\,155528, and found its value to be within $0.23^\circ$
of that cataloged by Schmidt et al. (1992a).  To check for instrumental
polarization, we observed the null standard BD $+32^\circ3739$, and found it to
be null to within $0.07\%$.  

All data obtained from the Lick, Arizona, and Palomar observatories
were extracted and calibrated using generic {\sc iraf}\footnote{IRAF
  is distributed by the National Optical Astronomy Observatory, which
  is operated by the Association of Universities for Research in
  Astronomy (AURA) under a cooperative agreement with the National
  Science Foundation (NSF).} routines and our own {\sc idl}
functions. Spectropolarimetric analysis was also performed in {\sc
  iraf} and {\sc idl} following the methods outlined by Miller,
Robinson, \& Goodrich (1988) and detailed by Leonard et al. (2001) and
Leonard \& Filippenko (2001). All data were corrected for the redshift
of M51 ($z=0.00155$).

\section{Analysis \& Results}
Linear polarization is expressed as the quadratic sum of the $Q$ and $U$ Stokes parameters, $P= \sqrt{q^2 + u^2}$, and the position angle on the sky is given by $\theta=(1/2)\,\textrm{tan}^{-1}(u/q)$, while carefully taking into account the quadrants in the $q$--$u$ plane where the inverse tangent angle is located.  Since $P$ is a positive-definite quantity, it is significantly overestimated in situations where the signal-to-noise ratio (S/N) is low.  It is typical to express the ``debiased" (or ``bias-corrected") form of $P$ as $P_{\rm db}=\sqrt{(q^2 + u^2) - (\sigma_{q}^2 + \sigma_{u}^2)}$, where $\sigma_q$ and $\sigma_u$ are the uncertainties in the $q$ and $u$ Stokes parameters.  If $(\sigma_{q}^2 + \sigma_{u}^2) > (q^2 + u^2)$, then we set a 1$\sigma$ upper limit on $P$ of  $\sqrt{\sigma_{q}^2 + \sigma_{u}^2}$. In cases where $P/\sigma_{P} < 1.5$, $\theta$ is essentially undetermined.  All polarized spectra presented herein are displayed in this manner. Note, however, that at low S/N, $P_{\rm db}$ is also not a reliable function, as it has a peculiar probability distribution (Miller et al. 1988). Thus, for extracting statistically reliable values of polarization within a particular waveband, we have binned the calibrated $q$ and $u$ Stokes spectra separately over the wavelength range of interest before calculating $P$ and $\theta$. All quoted and tabulated values in this paper were determined in this manner, although spectra are displayed as $P_{\rm db}$, so they may exhibit slight offsets from our quoted values.

   \begin{figure*}
\includegraphics[width=4.7in]{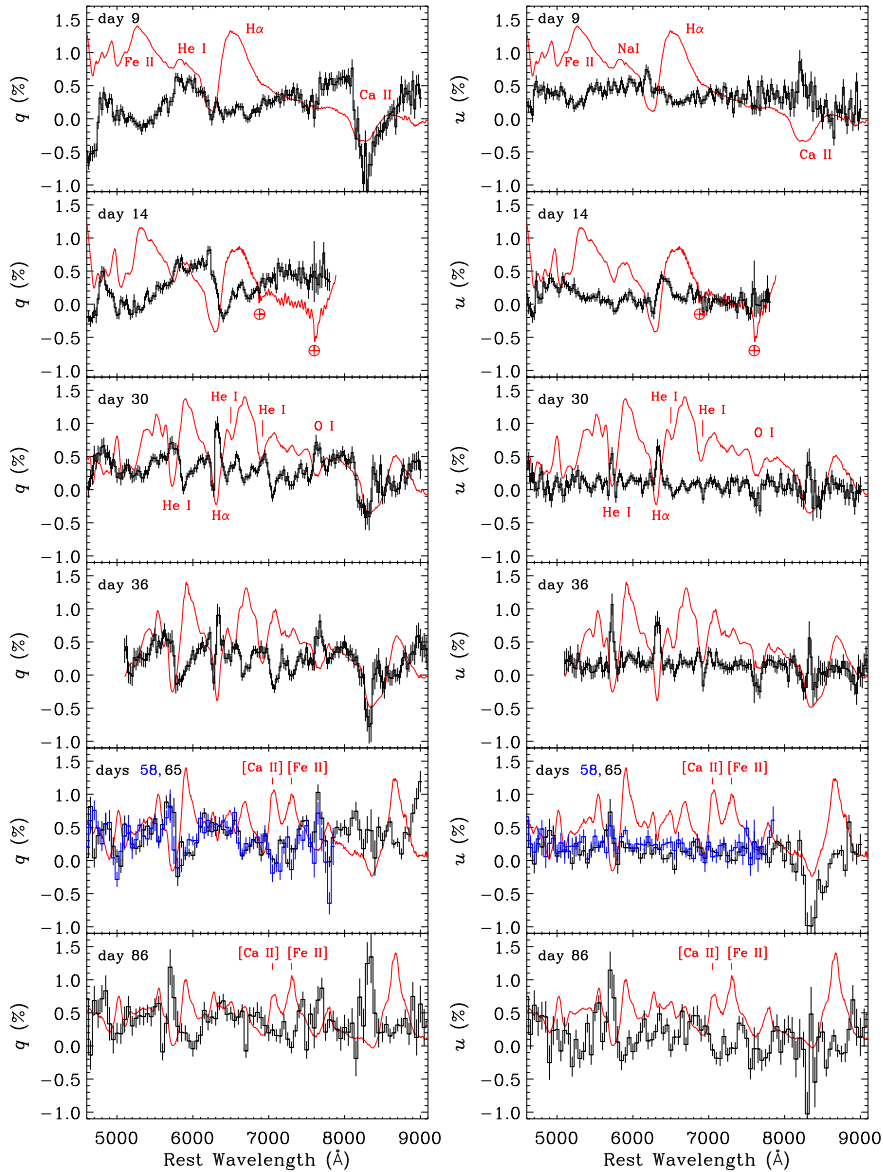}
\caption{Temporal evolution of the $q$ and $u$ Stokes parameters (black/blue curves) and total-flux spectra (red curves) of SN\,2011dh. The data are binned to 25\,{\AA}, except days 58 and later, which are binned to 50\,{\AA}. Prominent atomic features are labeled at different epochs. Telluric absorption lines in the day-14 data are marked with Earth symbols (circled crosses).}
\label{fig:qu}
\end{figure*}

 \begin{figure*}
\includegraphics[width=4.7in]{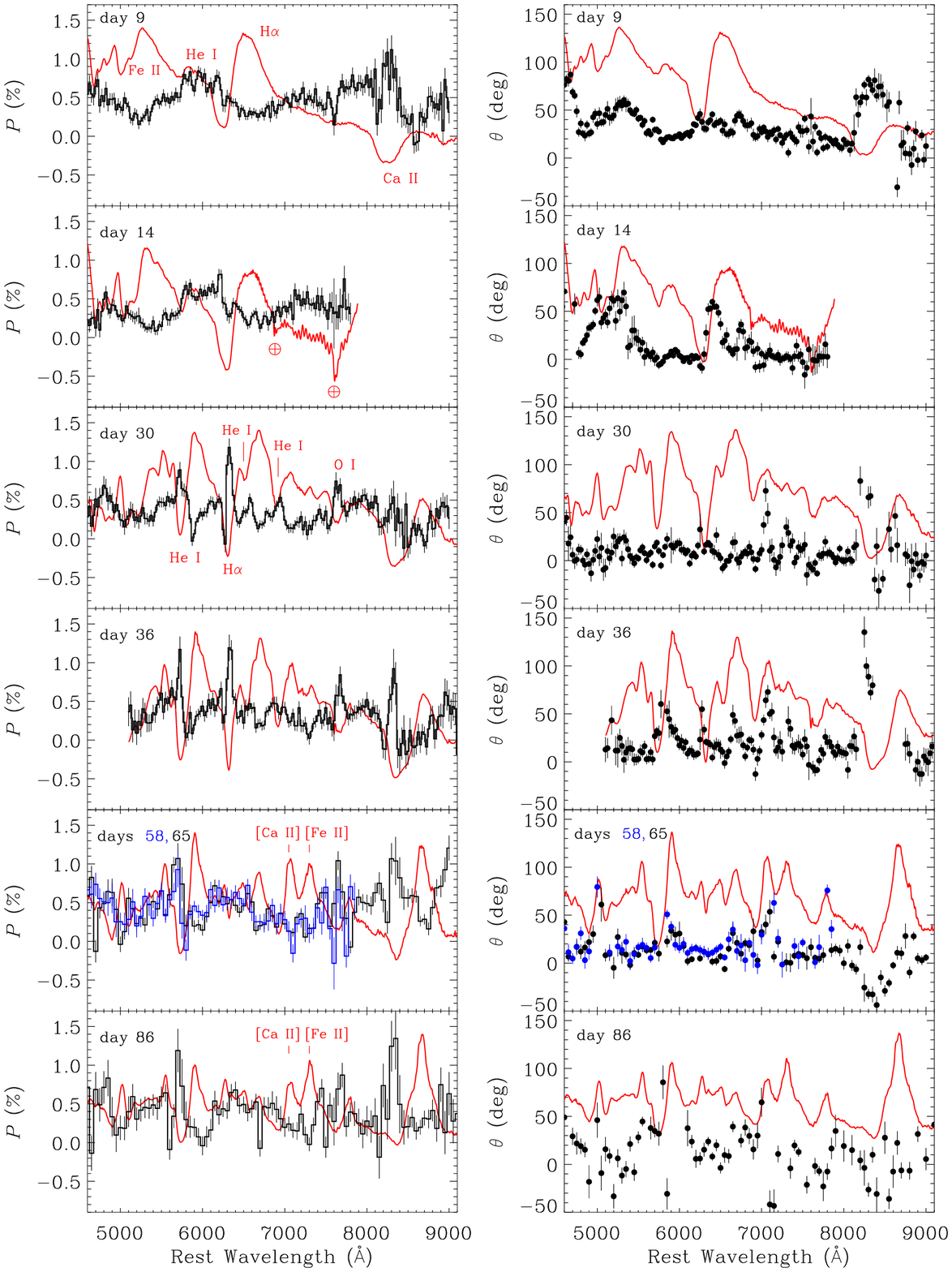}
\caption{Same as Figure\,\ref{fig:qu}, but for debiased polarization, $P$, and position angle, $\theta$. Wavelengths for which $P/\sigma_{P}<1.5$ have been omitted from the $\theta$ curve. Telluric absorption lines in the day-14 data are marked with Earth symbols (circled crosses).}
\label{fig:p_theta}
\end{figure*}

\subsection{Weak interstellar polarization}
As light traverses interstellar media, nonspherical dust grains aligned with the local magnetic field can polarize incident photons (Hiltner 1949; Davis 1955). Correcting for this component of interstellar polarization (ISP) is often a very problematic element of polarimetric analysis, as an accurate probe of the ISP from both the Milky Way and, in particular, the SN host galaxy, can be very difficult to obtain. According to Serkowski et al. (1975), the colour excess from interstellar extinction can be used to derive an upper limit on Galactic ISP, given by $P_{\rm ISP}<9E(B-V)$\%. The nondetection of Na\,D lines at the redshift of M\,51 by Arcavi et al. (2011) and Vink{\'o} et al. (2012) indicates that extinction from the host galaxy is low, with an estimated upper limit of $E(B-V)<0.05$\,mag (Arcavi et al. 2011). The Galactic component of extinction in the direction of SN\,2011dh is $E(B-V)=0.035$\,mag (Schlegel et al. 1998), for which we expect a maximum possible ISP of $<0.3$\%, based on the upper limit from Serkowski et al.

To directly gauge the effect of {\it Galactic} ISP, we observed the A5\,V star HD\,117815, which lies $<1^{\circ}$ away from SN\,2011dh on the sky and is presumed to be intrinsically unpolarized. It has a measured parallax that indicates a distance of 122\,pc (Perryman et al. 1997). At a Galactic latitude of $+68^{\circ}$, this corresponds to a Galactic scale height of 113\,pc, so the light from this star should probe the majority of intervening dusty ISM from the Galaxy. We observed this star with the Kast instrument on two occasions and detected no significant polarization, measuring only upper limits of $P_V<0.06\%$ on 2011 June 30 and  $P_V<0.03\%$ on 2011 July 6. We also observed the F5\,V star HIP\,65768 with Kast on 2016 Jun 30. The parallax of this object indicates a distance of 211\,pc and a Galactic scale height of 154\,pc. We measure $q_V=0.04\pm0.02$\% and $u_V=0.09\pm0.02$\%, which indicates $P=0.10\pm0.03$\%. At $P<0.1$\%, it appears that the {\it Galactic} component of ISP is practically insignificant in the direction of SN\,2011dh, and much lower than the maximum possible value of ISP indicated by the extinction.

To estimate the vector sum of ISP values from M51 and the Galaxy, one might proceed by assuming that the emission components of the strongest features, such as H$\alpha$, are completely depolarized, intrinsically (as was assumed in the case of SN\,1993J;  Tran et al. 1997). However, Maund et al. (2007a) demonstrated that using the polarization of H$\alpha$ emission for this purpose is probably not reliable, since there is the possibility for subtle blending of other features (e.g., He\,{\sc i} $\lambda$6678). Furthermore, if a SN has more than one component of polarization (for example, Thomson scattering and/or dust scattering by CSM), then this could produce net polarization at the same wavelengths as the strong intrinsically unpolarized emission features (e.g., see Mauerhan et al. 2014).  The possible existence of  extended CSM has been suggested in the case of SN\,2011dh, based on detection of broad H$\alpha$ in late-time nebular spectroscopy at $>200$ days (Sahu et al. 2013; Shivvers et al. 2013) and radio observations (Horesh et al. 2013). Whether this CSM component could be influencing the polarization properties of SN\,2011dh at the earlier times we are dealing with is not clear, but perhaps unlikely, given the lack of substantial UV excess reported for this object (e.g., see Ben-Ami et al. 2015). Nonetheless, the possibility of scattering by CSM warrants caution in assuming intrinsically depolarized emission features. Using the Ca\,{\sc ii} near-infrared (IR) triplet to estimate the total ISP might be a safer approach, since this region is devoid of other spectral features. The behaviour of this line makes a strong case for very low ISP, since at every epoch, $q$, $u$, and $P$ all trend close to null ($\lesssim0.1$\%) near the wavelength of the emission peak.

Another method for constraining the total vector sum of ISP from the host and the Galaxy is using late-time measurements of the Stokes parameters at the wavelengths of any strong forbidden lines as the SN enters the nebular phase. Such lines form in low-density regions of the explosion, where the effect of electron scattering is expected to be weak and the polarization should be insignificant. As will be demonstrated in the next section, the nebular-phase spectra indeed indicate a low value of ISP. We conclude that the total vector sum of ISP from the host and the Galaxy in the direction of SN\,2011dh is not substantial enough to make a significant impact on our measurements or scientific interpretation, and we thus move onward, making no attempt to remove ISP from the data.

\subsection{Spectropolarimetric evolution of SN 2011dh}
Figure\,\ref{fig:qu} illustrates the evolution of the $q$ and $u$ Stokes parameters of SN\,2011dh between 9 and 86 days post explosion, and Figure\,\ref{fig:p_theta} shows the same evolution for debiased polarization $P$ and position angle $\theta$. Each panel in both figures is accompanied by the total-flux spectrum at the given epoch. Table\,1 lists the integrated $V$-band and continuum values of $P$ and $\theta$ for each epoch. The most reliable sample of the SN continuum appears to lie in the wavelength range 7000--7450\,{\AA}.

\begin{figure*}
\includegraphics[width=2.5in]{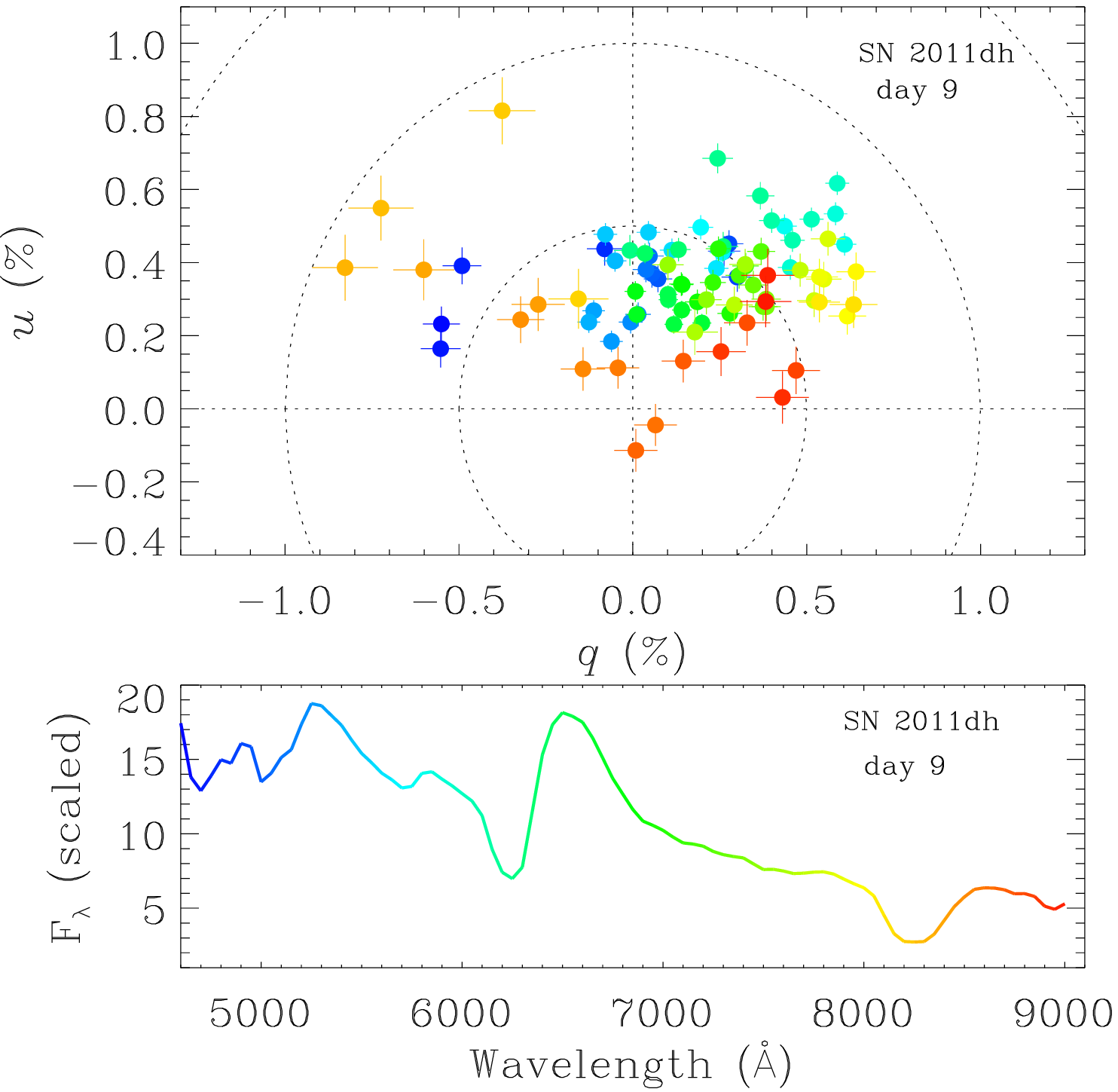}
\includegraphics[width=2.5in]{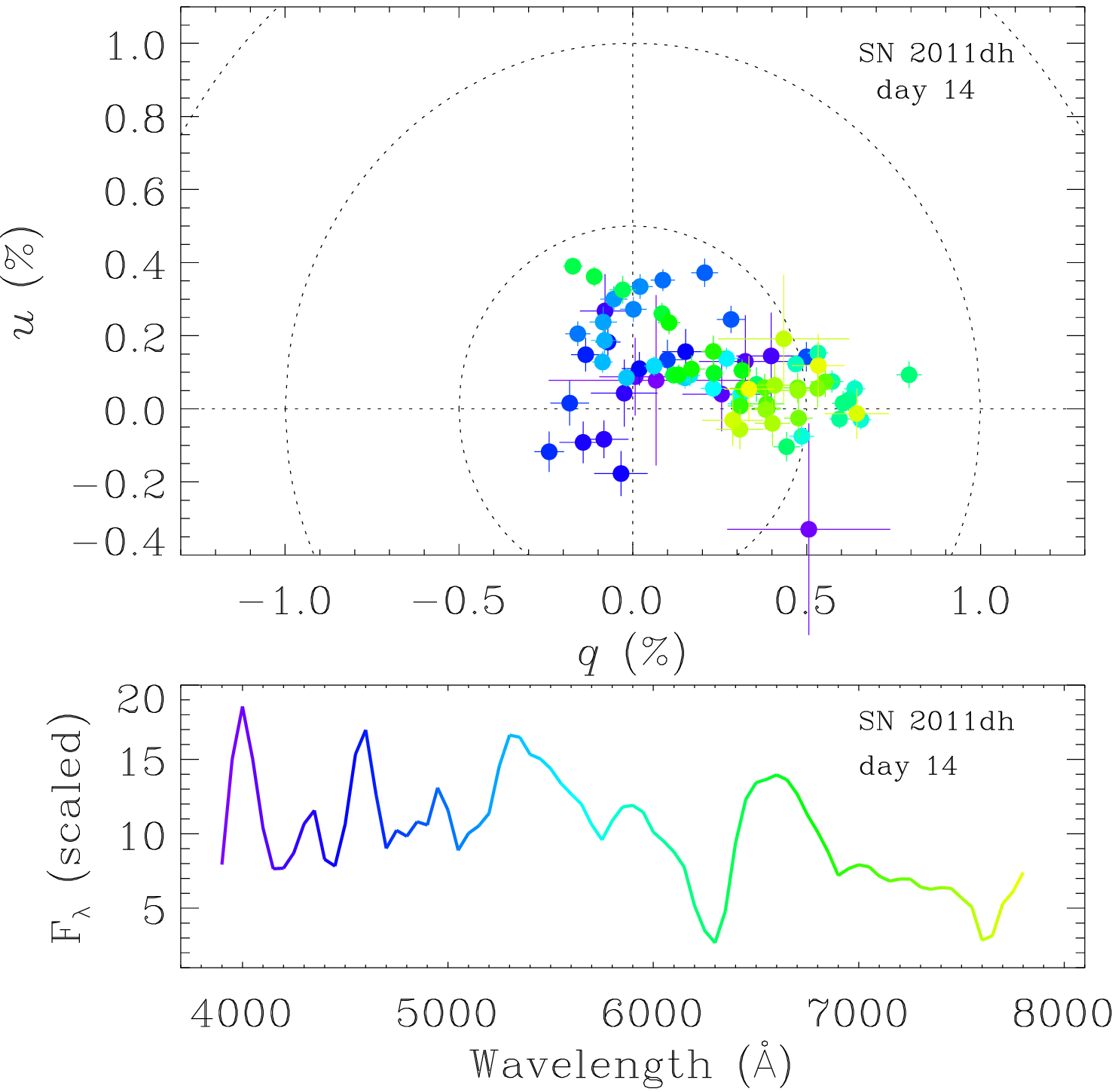}
\includegraphics[width=2.5in]{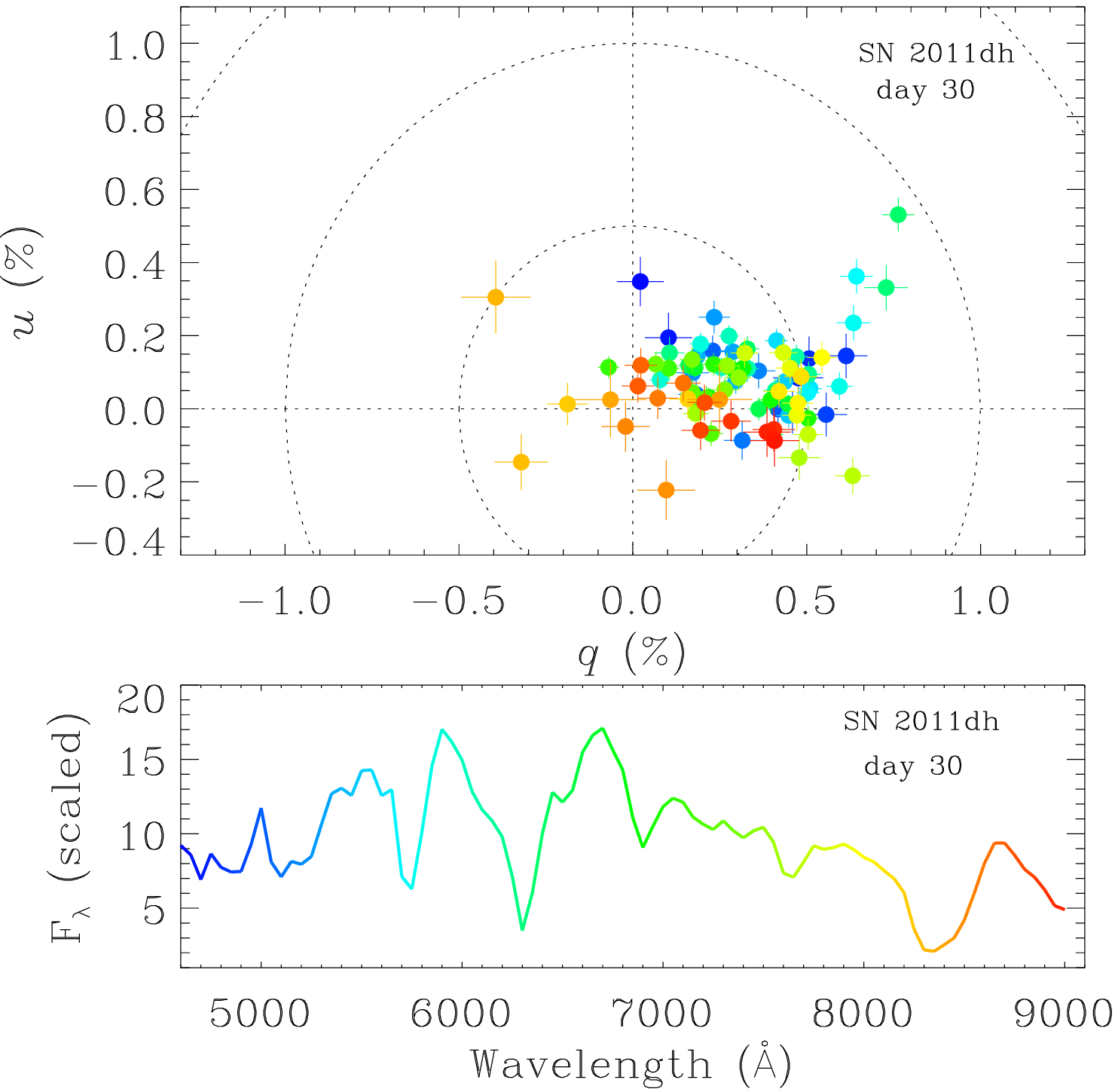}
\includegraphics[width=2.5in]{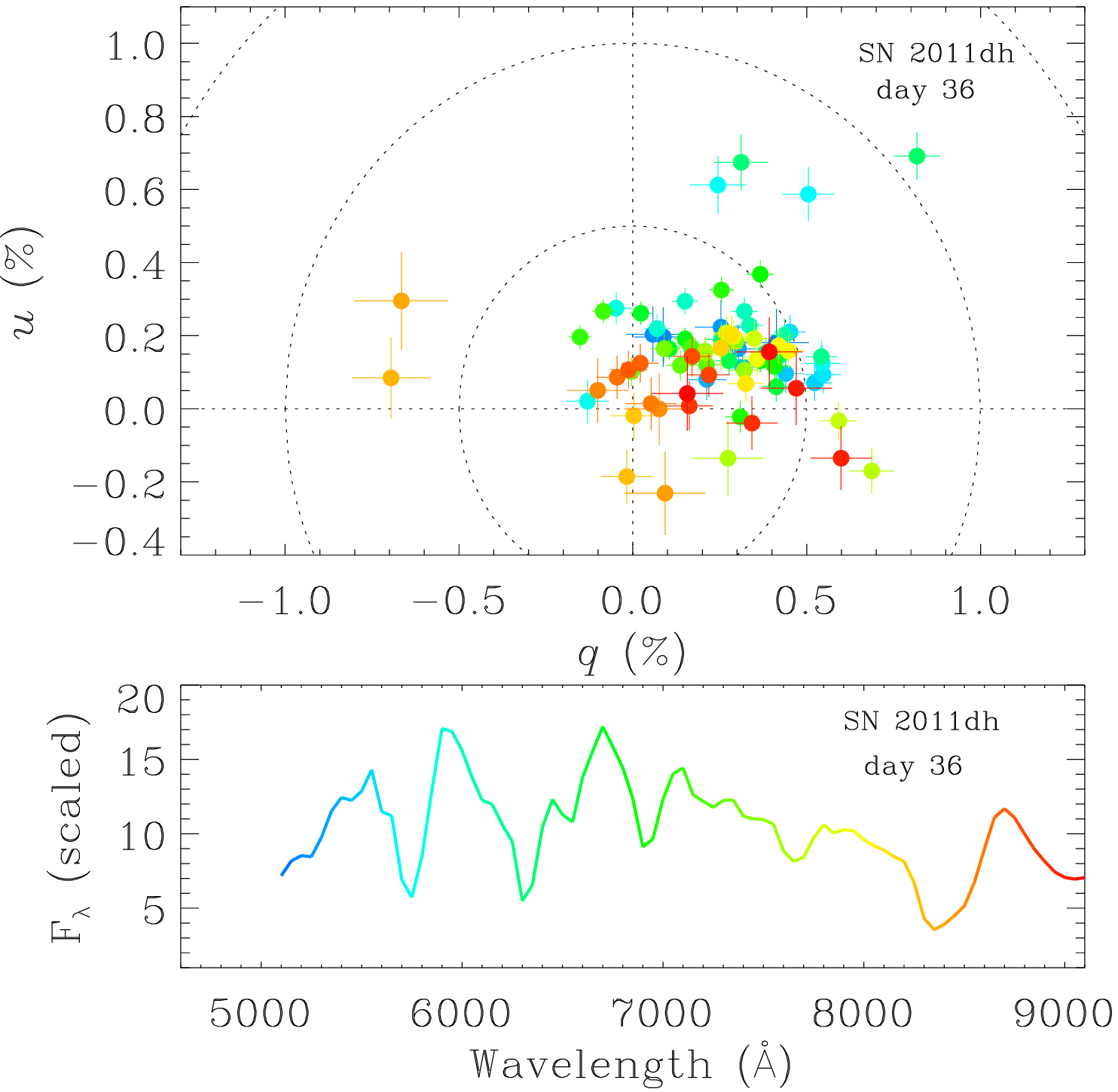}
\caption{Evolution in the $q$--$u$ plane of the entire spectrum on days 9, 14, 30, and 36. The data have been binned to 50\,{\AA}. Colours are used to indicate different parts of the spectrum. Note the smaller wavelength coverage on day 14. }
\label{fig:qu_all}
\end{figure*}

 \begin{figure*}
\includegraphics[width=6.8in]{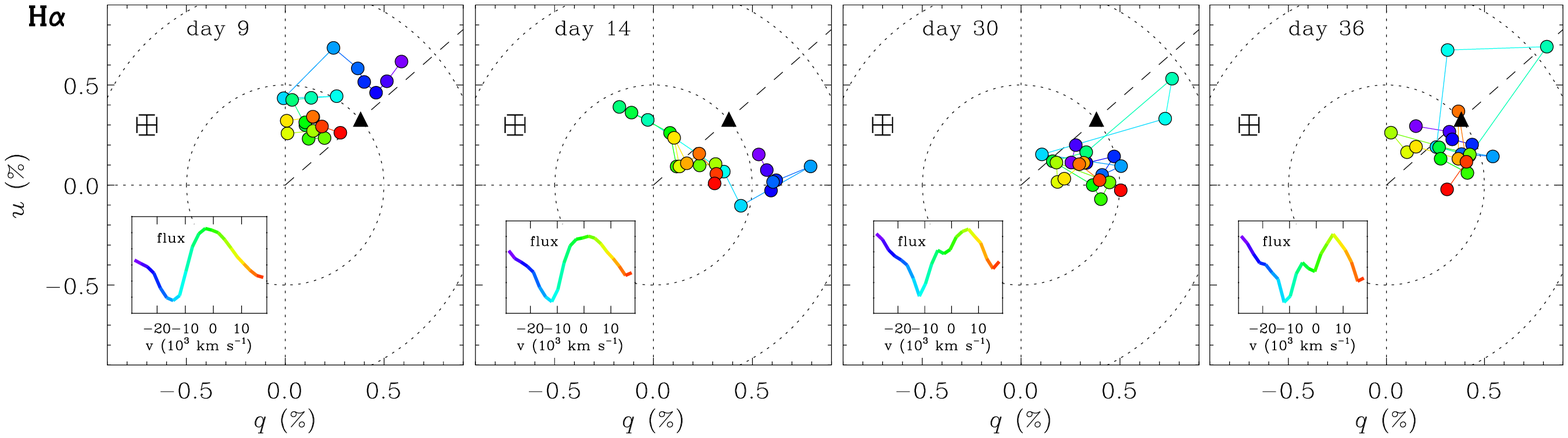}
\includegraphics[width=6.8in]{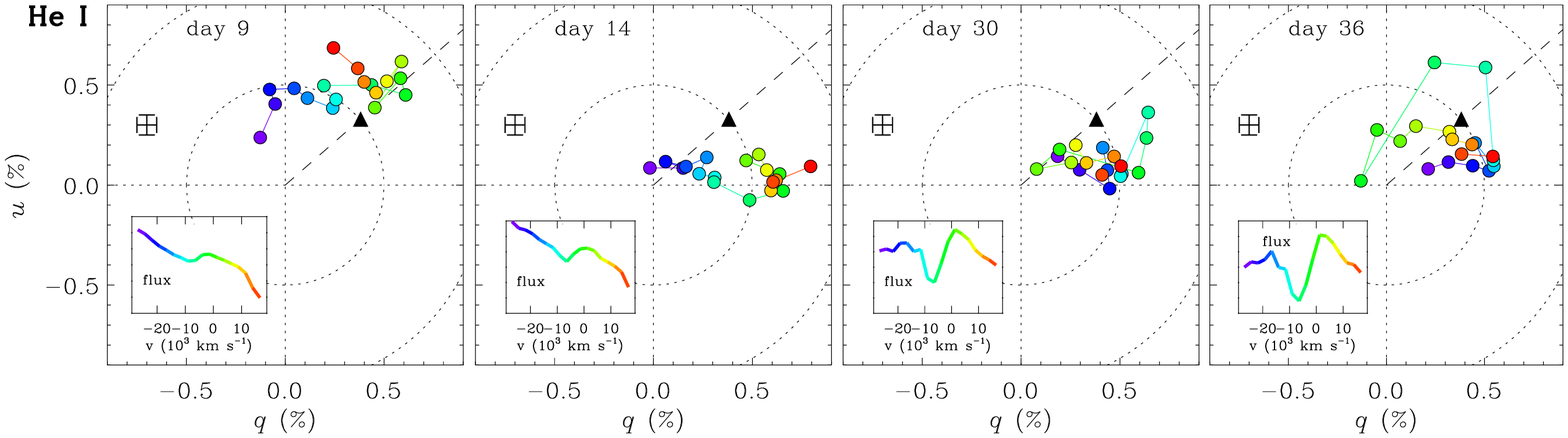}
\includegraphics[width=5.2in]{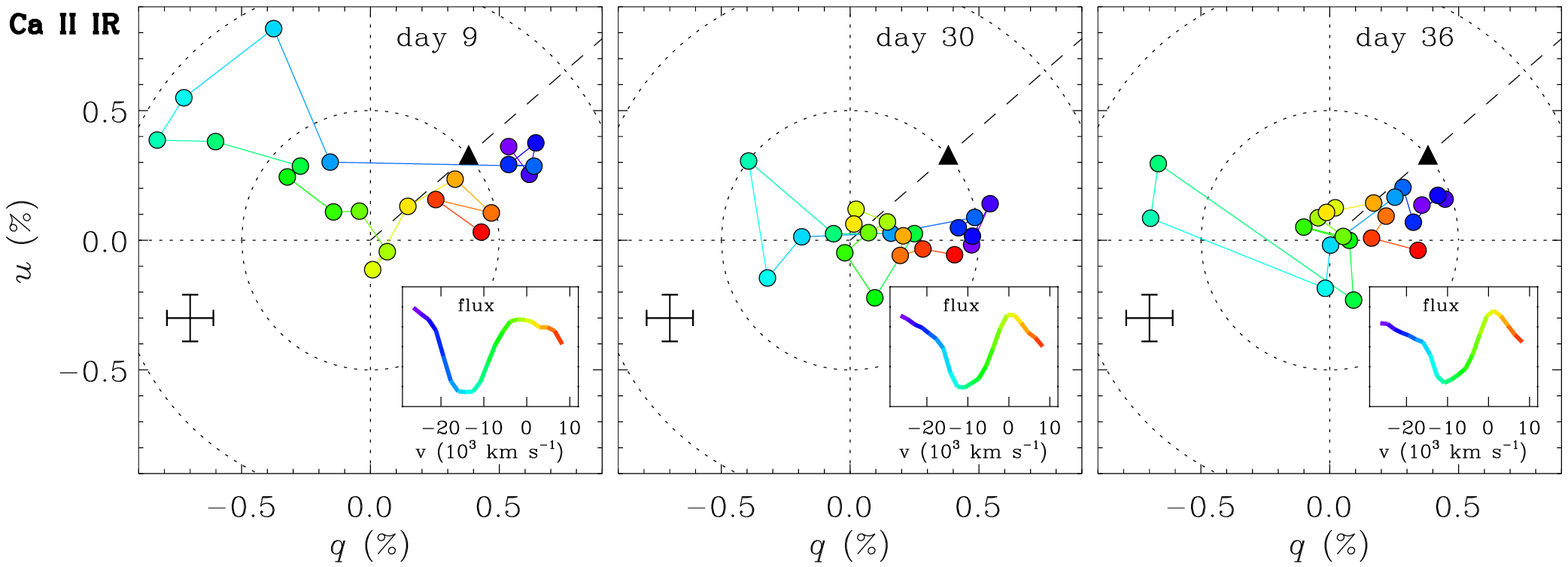}
\includegraphics[width=5.2in]{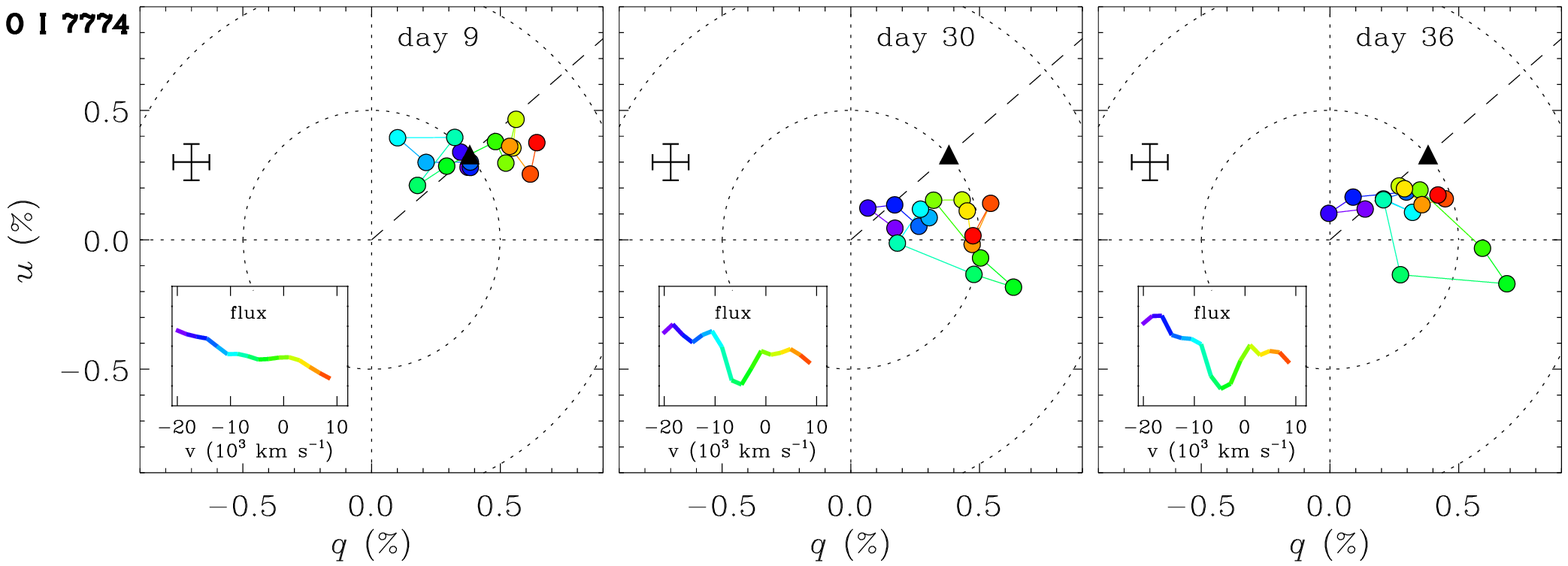}
\caption{Evolution in the $q$--$u$ plane for prominent spectral lines, colour coded by radial velocity. Four epochs are shown for H$\alpha$ and He\,{\sc i} $\lambda$5876, and three epochs are shown for the Ca\,{\sc ii} IR triplet and O\,{\sc i}. The inset frames display the flux spectra with arbitrary scaling. The dashed line represents the ``axis" defined by the day-9 continuum (black triangle). The Stokes data are binned to 50\,{\AA}. The black error bar in each panel represents the average uncertainty.}
\label{fig:qu_lines}
\end{figure*}

The earliest (day 9) data were obtained when the SN photosphere was still in the H envelope, as indicated by the Balmer-dominated spectrum. At this stage, the spectrum consists of broad P-Cygni lines from H$\alpha$, a blend of Fe\,{\sc ii} emission centered near 5300\,{\AA} with multiple emission/absorption components extending blueward of 5100\,{\AA}, and a strong feature from the Ca\,{\sc ii} IR triplet near 8650\,{\AA}. A relatively weak feature of Na\,{\sc i} or He\,{\sc i} is also seen in the range 5800--5900\,{\AA}, which for a SN\,IIb is most likely attributable to He\,{\sc i} $\lambda$5876 (see Hachinger et al. 2012), and thus we assume this identification hereafter. At this stage substantial polarization is detected. As is commonly seen in CC~SNe, troughs and crests in $P$ are coincident with the respective emission and absorption components of the P-Cygni profiles (``inverted P-Cygni" polarization profiles), likely resulting from asymmetric or patchy obscuration of the photosphere. Regions of broad polarization minima are associated with the broad emission components in the total-flux spectra, whereby the strong intrinsically unpolarized line emission dilutes the polarized flux (Cropper et al. 1988). Fluctuations of $\sim0.5$\% in $P$ are evident around He\,{\sc i} and H$\alpha$, and reach $\sim1$\% for Ca\,{\sc ii} IR.  A region of line-free continuum can be seen in the range 7000--7450\,{\AA}, where we measure $P=0.47\pm0.03$\% and $\theta=23.5\pm1.5^{\circ}$; continuum polarization also probably contributes to the broad hump in $P$ centered near 5000\,{\AA} on days 9 and 14, as indicated by the level of $P$ at the continuum near 6050\,{\AA}. The highest value of $P$ is associated with the absorption components of the Ca\,{\sc ii} IR triplet, reaching $\sim1.1$\%. Substantial rotations in $\theta$ are seen across the spectrum, changing by $\sim30^{\circ}$ across the Fe\,{\sc ii} blend, and fluctuating by $\sim20^{\circ}$ across H$\alpha$. The largest rotations are 80--90$^{\circ}$ across Ca\,{\sc ii} IR and $\sim70^{\circ}$ at the far blue end of the spectrum, possibly associated with H$\beta$. 

By day 14, He\,{\sc i} features are beginning to appear in the flux spectrum. The $\lambda$5876 line exhibits deepening absorption, and the $\lambda$6678 line can be seen as a notch of absorption near the peak of the H$\alpha$ emission component. Spectral analysis by Marion et al. (2014) showed that the first definitive detection of He\,{\sc i} lines occurred 11 days post explosion, so at day 14 the photosphere must be entering the He layer of the envelope. At this stage, the overall characteristics of $P$ are similar to those on day 9, although the broad enhancement near 6000\,{\AA} has shifted slightly toward the red, with the addition of a narrow peak of enhanced $P$ appearing near 6200\,{\AA}. Substantial evolution in $\theta$ has occurred between days 9 and 14, as the major rotations seen across the spectrum have grown in strength, most notably near the blueshifted portion H$\alpha$ (on the lower-velocity slope of the absorption trough), and also on the red shoulder of the emission component. The red end of our wavelength coverage at this epoch cuts off at 7800\,\AA, so the evolution of Ca\,{\sc ii} IR cannot be ascertained. It should be noted, however, that the large rotation in $\theta$ between days 9 and 14 appears to be dominated by the evolution of the continuum component. The 7000--7450\,{\AA} continuum region on day 14 exhibits $P=0.45\pm0.02$\% and $\theta=2.2\pm1.4^{\circ}$, comparable in strength to that on day 9 but shifted by $\Delta\theta\approx20^{\circ}$.

By day 30, the character of the total-flux spectrum has changed considerably. He\,{\sc i} lines have become dominant, indicating that the photosphere has moved into the He layer of the outflow. The emission profiles of H$\alpha$ and He\,{\sc i} $\lambda$5876 appear to be asymmetric, having a bump or knee between the peak wavelength and the red side of the profile. Enhanced $P$ is very prominent for the absorption component of H$\alpha$, and it exhibits a peculiar fluctuation in strength between the blue and red walls of the absorption trough. Specifically, $P$ reaches a maximum of $\sim1.2$\% on the red wall of the absorption trough but rapidly drops to null on the bluer, higher-velocity side of the trough. Curiously, the associated change in $\theta$ across this feature is unremarkable. $P$ drops to almost null values near the peaks of He\,{\sc i} and Ca\,{\sc ii} IR emission. Less-extreme enhancements in $P$ are also associated with the red walls of the absorption components for the various He\,{\sc i} lines throughout the spectrum. A new feature of enhanced $P$ is also seen for the absorption component of O\,{\sc i} near 7600--7700\,{\AA}. 

The most prominent rotation is a $\Delta\theta\approx90^{\circ}$ shift associated with Ca\,{\sc ii} IR. There is also a $\sim70^{\circ}$ rotation associated with a feature near 7000--7100\,{\AA}, which might be attributable to He\,{\sc i} $\lambda$7065, although in this case it would be unclear why similar rotations are not also seen for other He\,{\sc i} features. The continuum region used for days 9 and 14 appears to have dropped to $P = 0.18\pm0.04$\% and rotated to $\theta = 12.0^{\circ}\pm3.4^{\circ}$, but this could be heavily influenced by the new appearance of line structure. On the other hand, the broad polarization hump centered near 5000\,{\AA} on days 9--14, which is continuum polarization, has dropped as well. It thus appears that the continuum polarization decrease between days 14 and 30 could be real and substantial.

The day-36 spectrum exhibits only slight changes in H$\alpha$ and He\,{\sc i} polarization with respect to day 30, and the emission profiles still appear to exhibit the asymmetric knee on the red sides.  However, substantial changes in $P$ and $\theta$ are evident for other lines. In particular, we see strong enhancement in $P$ for the highest-velocity absorption of Ca\,{\sc ii} IR. Enhanced $P$ associated with He\,{\sc i} $\lambda$5876 and O\,{\sc i} have also grown to $\sim1$\%.

By day 65, the absorption component of H$\alpha$ is almost completely gone, implying that the H-rich outer layers of the outflow are close to becoming fully transparent. P-Cygni profiles of He\,{\sc i} $\lambda$5876, Ca\,{\sc ii} IR, and O\,{\sc i} are still prominent, however, each maintaining enhanced $P$ levels of $\sim1$\% for their blueshifted absorption components. Forbidden emission lines of [Fe\,{\sc ii}] and [Ca\,{\sc ii}] have appeared in the wavelength range 7000--7400\,{\AA}, heralding the nebular transition. By day 86, the P-Cygni absorption components of all lines have weakened as the explosion thins out, but strongly enhanced $P$ is still seen for blueshifted He\,{\sc i}  $\lambda$5876 and for Ca\,{\sc ii} IR, in addition to weaker yet significant features across the rest of the spectrum.  We note that the asymmetric knee on the red side of the He\,{\sc i} $\lambda5876$ profile, which first became apparent on day 30, is still present through day 86.

Since the forbidden lines seen at late times form in low density regions of the outflow, electron scattering is weak, and thus the polarization should be insignificant near the central wavelength of each line. As an additional means of estimating the ISP, we averaged the Stokes parameters over the regions corresponding to the peak fluxes of the lines (7020--7110\,{\AA} and 7270--7350\,{\AA}). We obtained $q_{\textrm{ISP}}=0.08\pm0.05$\% and $u_{\textrm{ISP}}=-0.01\pm0.05$\%, which implies that $P=0.08\pm0.07$\% at these wavelengths. The low value of ISP indicated by this result supports our decision in \S3.1 to make no attempt at removing ISP from the data. 

\begin{figure}
\includegraphics[width=3.in]{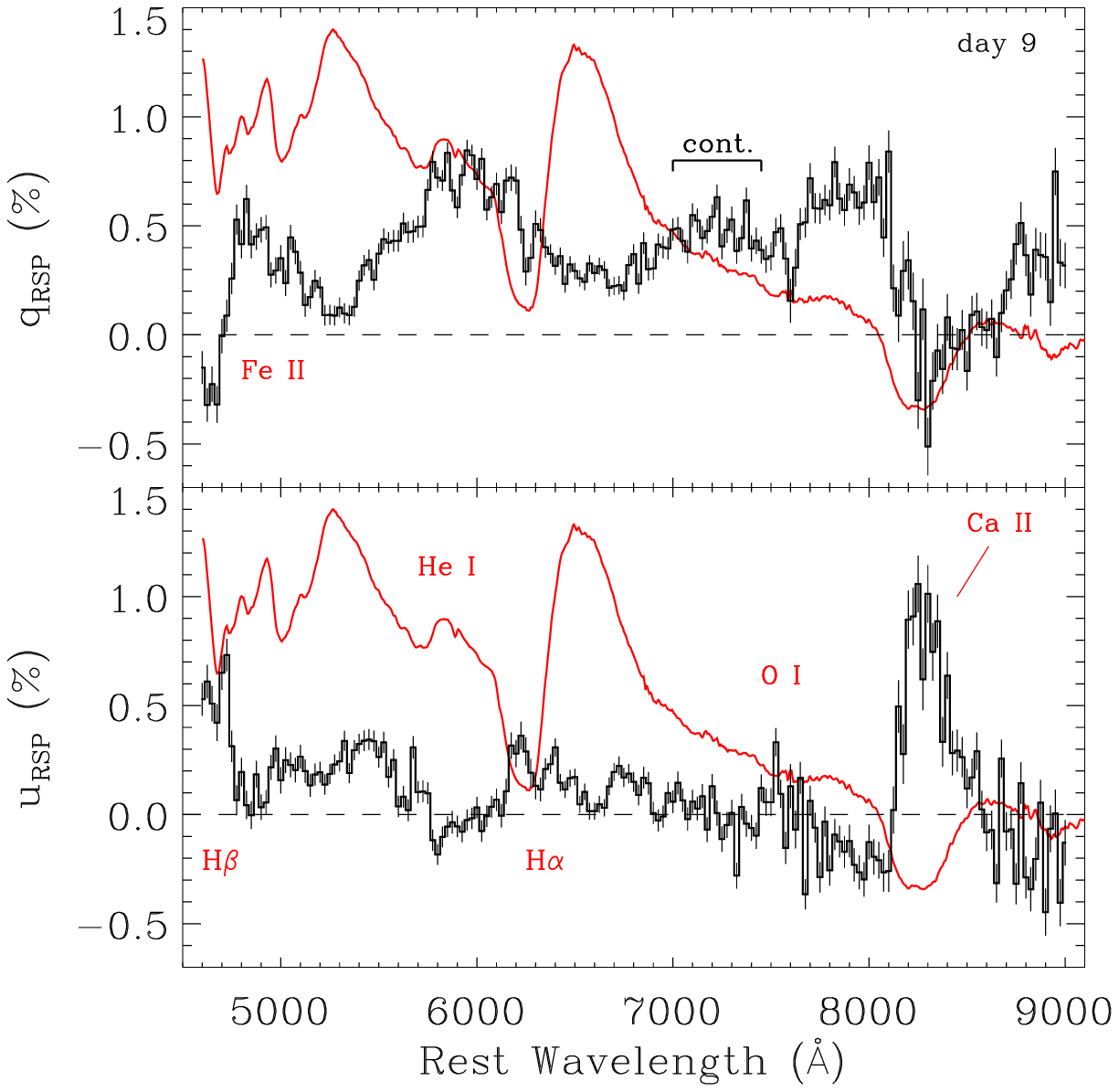}
\includegraphics[width=3.in]{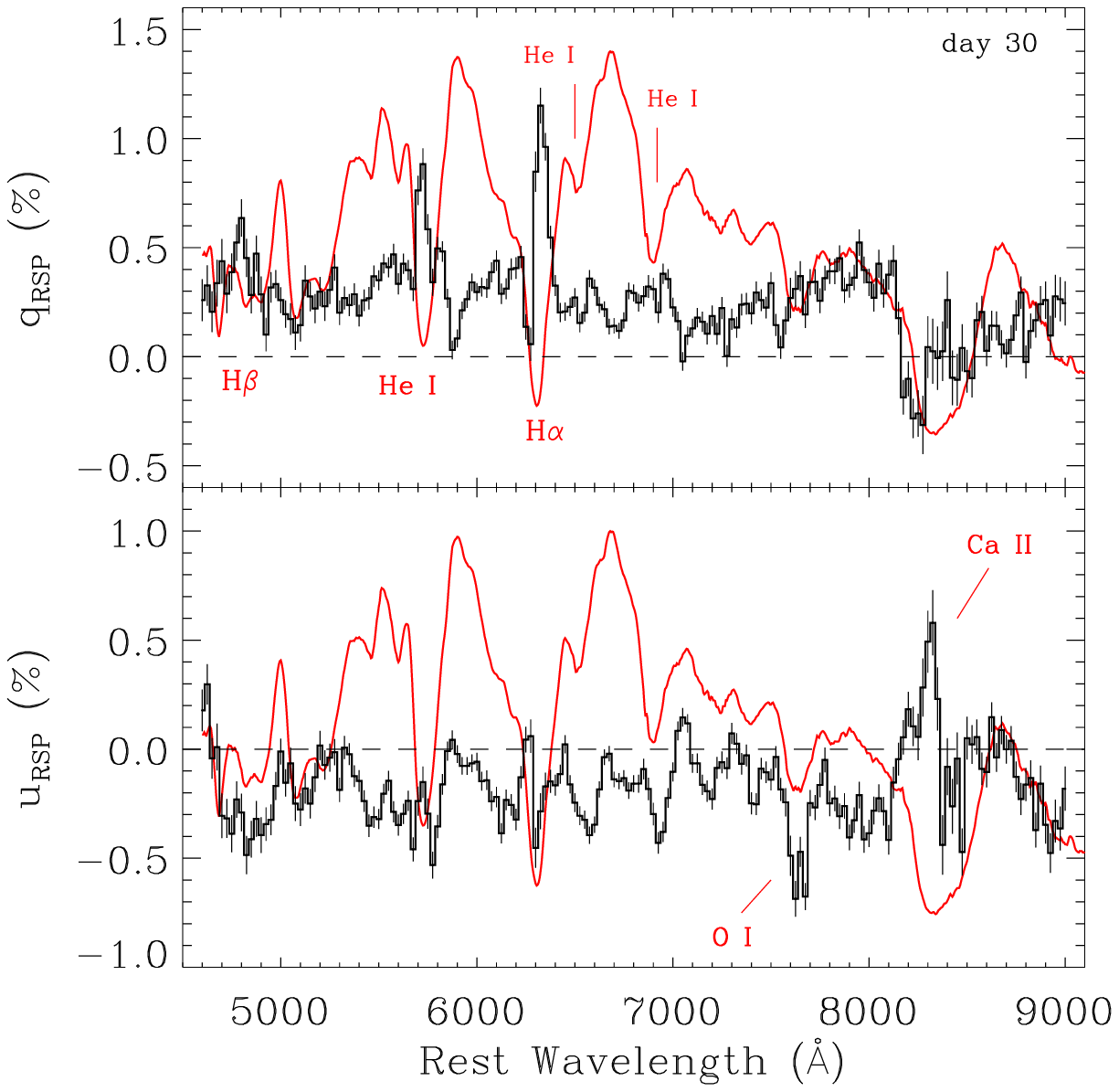}
\caption{Rotated stokes parameters for SN\,2011dh on days 9 and 30, aligned to the axis of symmetry defined by the day-9 continuum region (7000--7450\,{\AA}). Total-flux spectra are shown in red.}
\label{fig:rsp}
\end{figure}

\begin{figure}
\includegraphics[width=1.72in]{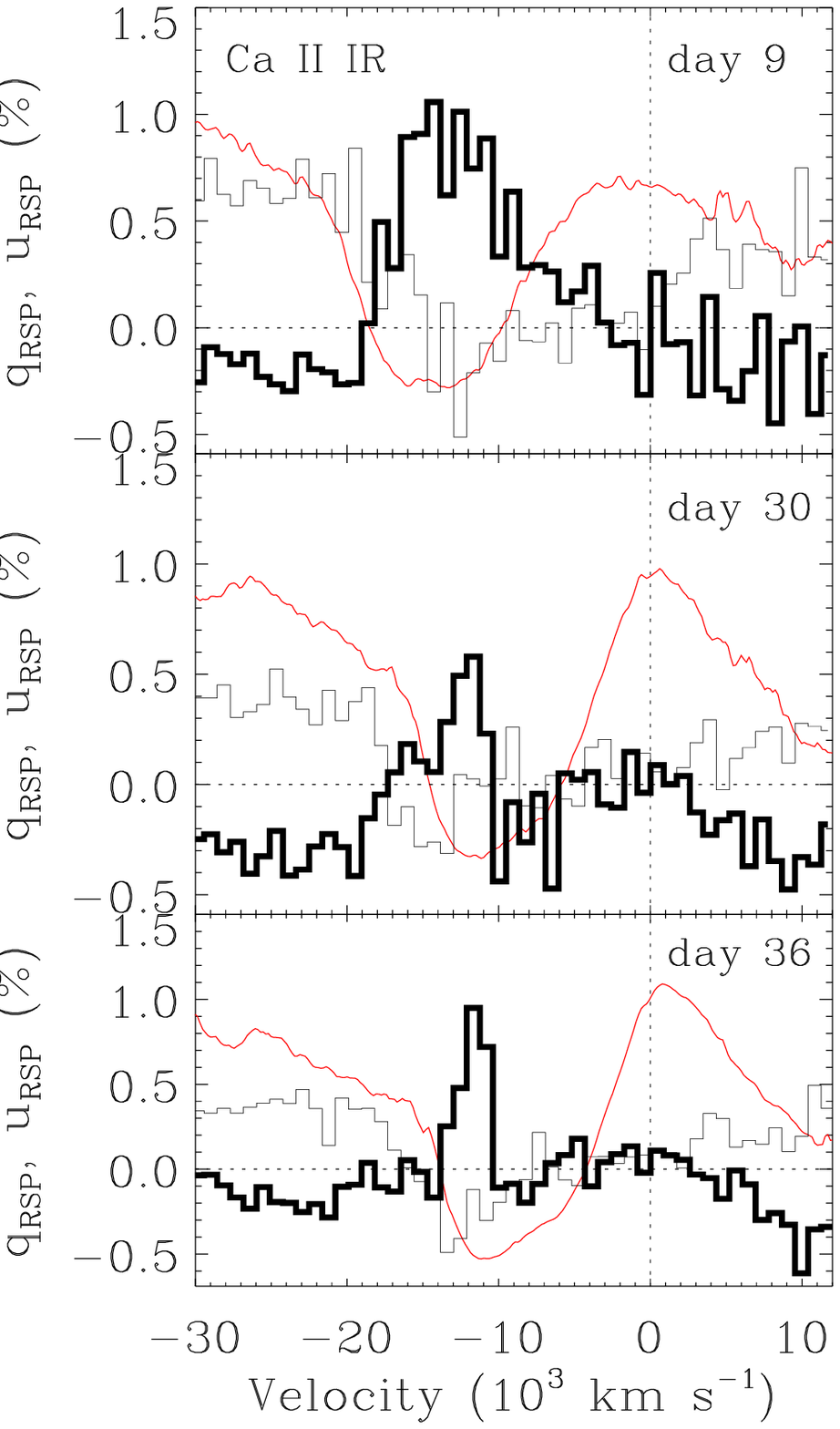}
\includegraphics[width=1.57in]{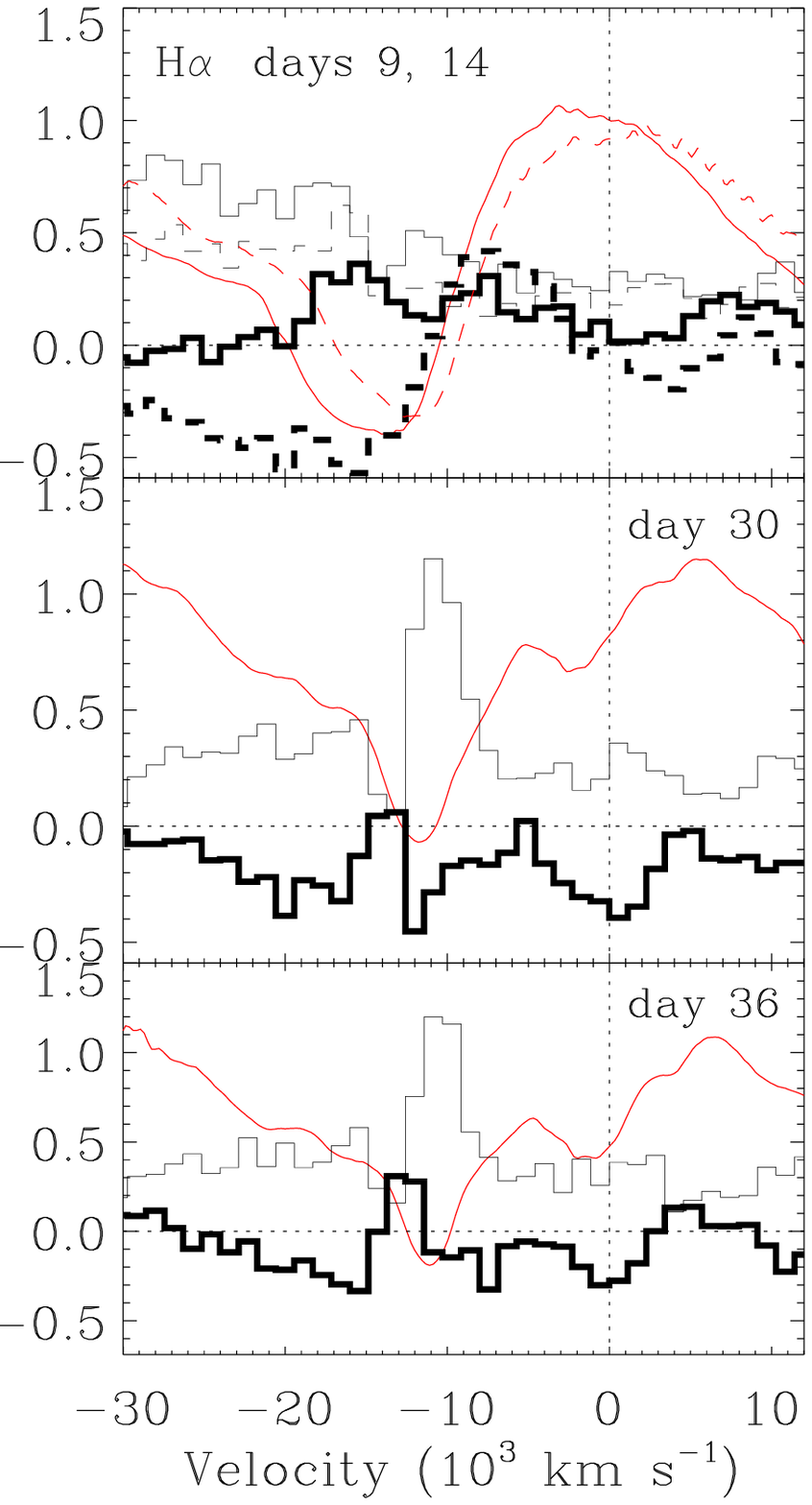}
\caption{Evolution of $q_{\textrm{RSP}}$ (thin black curves) and $u_{\textrm{RSP}}$ (thick black curves) as a function of radial velocity for the Ca\,{\sc ii} IR triplet and H$\alpha$. Total-flux spectra are shown in red. In the upper-right panel, the day-14 data are represented by the dashed curves.}
\label{fig:ca_polvel}
\end{figure}

\subsubsection{Behaviour in the $q$--$u$ plane}
Figure\,\ref{fig:qu_all} shows the $q$--$u$ plane evolution over the full wavelength range of the spectra for days 9, 14, 30, and 36. Figure\,\ref{fig:qu_lines} illustrates the behaviour in the $q$--$u$ plane across prominent line features, and also shows the flux spectra of the line plotted with respect to radial velocity (RV).  On day 9, the full range of data forms a cloud of points that are roughly centered in quadrant\,{\sc i}. The large excursion near the Ca\,{\sc ii} IR triplet is clearly seen as a looping pattern that moves far into quadrant\,{\sc ii}, with maximum excursion occurring for RVs ranging from $-10,000$\,km\,s$^{-1}$ to $-15,000$\,km\,s$^{-1}$.  This feature is mostly nonaxisymmetric relative to the 7000--7450\,{\AA} continuum sample (shown as a dashed line). However, the fastest Ca\,{\sc ii} absorption on day 9 has RV $=-20,000$\,km\,s$^{-1}$ and moves back into quadrant\,{\sc i}, nearer to the continuum axis. 

Blueshifted H$\alpha$ absorption creates a loop in the $q$--$u$ plane that, on day 9, deviates from the continuum axis for RV below $-20,000$\,km\,s$^{-1}$. On day 14, the large rotation of $\theta$ near H$\alpha$ is clearly apparent in the $q$--$u$ plane, but by days 30--36 the $q$--$u$ loops associated with this line, and with He\,{\sc i}, become more elongated and roughly aligned with the position angle of the day 9--14 continuum axes. The velocities at maximum polarization for H and He are $v\approx-10,000$\,km\,s$^{-1}$ and $v\approx-8000$\,km\,s$^{-1}$, respectively. The pronounced variation in $\theta$ observed across H$\alpha$ on day 14 could be the result of the evolving continuum polarization level; the changing vector addition of the Stokes parameters of this continuum component with those of H$\alpha$ could produce an artificial rotation across the line, rather than being, for example, an intrinsic change in the geometry of H$\alpha$ absorption itself.

The O\,{\sc i} feature, like Ca\,{\sc ii} IR, loops in a highly nonaxisymmetric manner with respect to H$\alpha$, He\,{\sc i}, and the day-9 continuum (most apparent on days 30--36), although the $q$--$u$ loop of this line drifts into quadrant {\sc iv} rather than quadrant {\sc ii}, as Ca\,{\sc ii} IR does. The velocity at maximum polarization for Ca\,{\sc ii} is comparable to that of H$\alpha$, while for O\,{\sc i} the velocity  is $\sim-4000$\,km\,s$^{-1}$, substantially slower than Ca\,{\sc ii}, H$\alpha$, and He\,{\sc i}.

To further illustrate the distribution of line features with respect to the day-9 continuum axis, we rotated the $q$ and $u$ Stokes parameters (RSP) to $\theta=23.5^{\circ}$; the results are shown in Figure\,\ref{fig:rsp}. It can be seen that the majority of polarization at wavelengths shortward of 8000\,{\AA} lies in $q_{\textrm{RSP}}$, along the axis of the continuum. The $q$--$u$ loop created by the Ca\,{\sc ii} feature is evident as a large deviation in the RSP, with most of the polarization contained in $u_{\textrm{RSP}}$ (nonaxisymmetric with the continuum). We examine this feature more closely in Figure\,\ref{fig:ca_polvel}, which illustrates $q_{\textrm{RSP}}$, $u_{\textrm{RSP}}$, and flux as a function of RV for Ca\,{\sc ii} and H$\alpha$. It can be seen that on day 9, the polarized profile of Ca\,{\sc ii} is relatively broad compared to later epochs. In the flux spectrum there appears to be two distinct components of absorption, with the most highly blueshifted component exhibiting an edge that extends to  RV $\approx-20,000$\,km\,s$^{-1}$, and the same is true for H$\alpha$. For Ca\,{\sc ii}, most of the polarized flux on day 9 is contained in $u_{\textrm{RSP}}$ for RV $\lesssim$ 15,000\,km\,s$^{-1}$, while at higher RV some of the polarization shifts into $q_{\textrm{RSP}}$. By day 36, the polarized feature exhibits a lower range of velocities centered at $\sim-12,000$\,km\,s$^{-1}$. Note that the emission features of Ca\,{\sc ii} and H$\alpha$ both exhibit a significant blueshift of 2000--3000\,km\,s$^{-1}$ on day 9, and appear to become more centered near rest velocity by days 14--30.

\section{Discussion}
\subsection{Sources of continuum and line polarization}
\subsubsection{The potential effect of stellar deformity}
The substantial continuum polarization detected from SN\,2011dh at the earliest epoch on day 9 implies that
the photosphere exhibited global asphericity while in the outer H-rich envelope. When interpreted in terms of the oblate, 
electron-scattering model  atmospheres of H{\"o}flich (1991), our results imply an asymmetry of at least 10\%; the actual 
value naturally depends on the (unknown) viewing orientation, and on the extension and depth of the electron-scattering atmosphere. The subsequent drop in continuum polarization to $P\approx0.2$\% by day 30, if real and not the result of line contamination starting at that epoch, suggests that the photosphere took on a more spherical configuration after passing into the He and C$+$O layers. This behaviour is opposite the polarimetric trends observed in some well-studied SNe\,II-P, which begin with low values of $P$ during the first few months, indicative of a mostly spherical outer envelope, and then increase in polarization only after H recombines and the photosphere enters the He and C$+$O layers; in these cases, the global asphericity has its origin in the core (Jeffery 1991; Leonard et al. 2001, 2006; Wang \& Wheeler 2008). 

For SN\,2011dh, a higher degree of global asphericity on the \textit{outside} possibly suggests that some other physical mechanism, perhaps related to binary evolution, might be at work. After all, in the majority of cases, SN\,IIb progenitors are thought to have lost the bulk of their H envelopes to tidal stripping by a massive accretor (Podsiadlowski et al. 1993; H{\"o}flich 1995; Smith et al. 2011; Dessart et al. 2012). Interestingly, a hot star has now been detected at the explosion site of SN\,2011dh, and it is a viable candidate companion to the progenitor (Folatelli et al. 2014). If the strong gravitational tides that induce mass transfer can also geometrically distort the envelope of the progenitor star, or synchronise the star's rotation at a fast enough rate to oblately deform it, then the star could develop an aspherical density profile. 

Interestingly, recent interferometric observations of the interacting yellow hypergiant binary HR\,5171A have provided evidence for significant envelope distortion in the primary (Chesneau et al. 2014). Although more luminous than the progenitor of SN\,2011dh, the example of HR\,5171A indeed motivates the expectation of binary-induced stellar distortion. If similarly induced geometric deformity in the progenitor of SN\,2011dh is substantial enough, the resulting aspherical density gradient could provide a means to generate an aspherical shock front and explosion geometry (Yamada \& Sato 1990; Steinmetz \& H{\"o}flich 1992). It should first be explored, however, whether sufficient tidal deformation is plausible for the SN\,2011dh progenitor system.

To assess the potential for tidal deformation, we require estimates of the physical parameters of the binary system and its component stars. For the case of SN\,2011dh, Benvenuto et al. (2013) performed binary stellar evolutionary calculations that followed the simultaneous evolution of both stars in the system to produce a pre-SN progenitor that is consistent with the space-based photometry (Van Dyk et al. 2011, 2013). Benvenuto et al. concluded that the most plausible model is one in which the primary and secondary components of the binary began with respective initial masses of $M_1^i=16\,{\rm M}_{\odot}$ and $M_2^i=10\,{\rm M}_{\odot}$, and an orbital period of $P=125$\,d. It had a moderate mass-transfer efficiency factor of $\beta=0.5$ and a specific angular momentum value of $\alpha=1$ (i.e., the angular momentum of the material lost by the system is equivalent to the angular momentum of the primary star). At the moment of explosion, the model predicted that the YSG primary would have a heavily reduced final mass of $M_1^f\approx4\,{\rm M}_{\odot}$, a small hydrogen content of $M_{\rm H}\approx3$--$5\times10^{-3}\,{\rm M}_{\odot}$, and a final radius of $R_{\textrm{\tiny{YSG}}}\approx250\,{\rm R}_{\odot}$ (consistent with the photometry); the mass-gaining secondary ended up with a final mass of $M_2^f\approx16\,{\rm M}_{\odot}$. Note that $\beta$ is one of most unconstrained parameters in binary evolution; regardless, considering the full range of possible values in their models ($0<\beta<1$), Benvenuto et al. concluded that the final semimajor axis of the system at the time of explosion lies in the range $A_f=842$--1009\,R$_{\odot}$, which implies $A_f/R_{\textrm{\tiny{YSG}}}\approx4$. Meanwhile, the ratio of the YSG's Roche-lobe radius and the final separation can be approximated by $R_{\textrm{\tiny{RL}}}/A_f\approx0.46\,M_1/(M_1+M_2)^{1/3}$ (Eggleton 1983), which in our case yields $R_{\textrm{\tiny{RL}}}/A_f\approx3.5$. Thus, $R_{\textrm{\tiny{YSG}}}\approx R_{\textrm{\tiny{RL}}}$.  

Frankowski \& Tylenda (2001) present analytic solutions for the geometric dimensions of a triaxial ellipsoid (TE) that stars will develop in response to tides in a massive binary (a football-like shape).  The principal axes of the ellipsoid are defined by the direction toward the companion (semi-axis $a_1$), the direction in the orbital plane that is perpendicular to the latter (semi-axis $a_2$), and the direction perpendicular to the orbital plane (semi-axis $a_3$), each given by 
\begin{equation}
a_1=R\left[1+ \left(\frac{1}{6\gamma}\right)\,(1+7q)\left(\frac{R}{A}\right)^3 \right], \nonumber
\end{equation}
\begin{equation}
a_2=R\left[1+ \left(\frac{1}{6\gamma}\right)\,(1-2q)\left(\frac{R}{A}\right)^3 \right],
\end{equation}
\begin{equation}
a_3=R\left[1- \left(\frac{1}{6\gamma}\right)\,(2+5q)\left(\frac{R}{A}\right)^3 \right], \nonumber
\end{equation}
where $q$ is the mass ratio, $R/A$ is the ratio of stellar radius to binary separation, and $\gamma$ is the ``gravity effectiveness" parameter, which is some value between zero and unity. For relatively compact main-sequence dwarfs, $\gamma \approx 1$ is an acceptable approximation that leads to only very subtle deformity for $R/A\approx1/4$. For YSGs, which are dynamically unstable (e.g., see Stothers \& Chin 2001), $\gamma$ could be $\sim0.1$ or lower, as a result of the relatively intense radiation pressure on the envelope, strong winds and high mass-loss rates, convection, and pulsations; such factors will profoundly influence the effective gravity and Roche potential, and can act to shrink the effective $R_{\textrm{\tiny{RL}}}$. If we therefore assume $\gamma=0.1$ and adopt the values of $q$ and $R/A$ above, Equation (9) from Frankowski \& Tylenda (2001) implies TE semi-axes of ($a_1$\,:\,$a_2$\,:\,$a_3$)\,$\approx\,$(1.1\,:\,1.0\,:\,0.9), or $\sim10$\% deviation from spherical symmetry along two axes. Lower values of $\gamma$, $q$, and $R$ will, of course, result in a higher degree of deformity, as will smaller values of $A$.

Stellar asphericity can also result from rotation, which might become fast enough to deform the star into an oblate geometry if the binary is tidally locked; this is a likely possibility for extended stars that nearly fill their Roche lobes. If so, the binary parameters adopted from Benvenuto et al. (2013) imply that the YSG would have a rotational velocity of $v_{\textrm{\tiny{YSG}}}\approx17$\,km\,s$^{-1}$. The critical rotational velocity, at which point the star will begin to break up, is given by $v_{\textrm{\tiny{crit}}}=\sqrt{GM_{\textrm{\tiny{eff}}}/R}$ (Langer et al. 1997; Langer 1998), where $M_{\textrm{\tiny{eff}}}=M(1-\Gamma_e)$ and $\Gamma_e\approx2.1\times10^{-5}\,L/M$. For the YSG progenitor, the values $L=10^{4.9}\,{\rm L}_\odot$  and $M_f=4\,{\rm M}_\odot$ imply $M_{\textrm{\tiny{eff}}}=3/4M_f=3\,{\rm M}_\odot$; thus, $v_{\textrm{\tiny{crit}}}=48$\,km\,s$^{-1}$ and $v_{\textrm{\tiny{YSG}}}/v_{\textrm{\tiny{crit}}} \approx 0.35$. This subcritical value would generate a rotationally induced asphericity of $\lesssim5$\%. Of course, this all depends on the model-dependent binary parameters we have adopted; a shorter binary separation or larger stellar masses would result in faster stellar rotation for a tidally locked system, and thereby larger asphericities.

A deformed stellar envelope, whether induced by tidal stretching, rotation, or both, will exhibit an aspherical density profile, having a steeper gradient along the poles; this can have a profound effect on the geometry of the photosphere of the subsequent SN explosion. Specifically, hydrodynamic simulations of shock propagation in rotating blue supergiant stars (like the progenitor of SN\,1987A) have demonstrated that an initially spherical shock wave that encounters a steeper density gradient along the poles will develop a prolate-shaped front and break out of the poles first (Steinmetz \& H{\"o}flich 1992). The interaction of this prolate shock with the oblate density stratification in the star causes a second shock wave that moves from the poles to the equator. The resulting transport of energy and momentum subsequently produces a quasihomologous outflow having an oblate outer and prolate inner density stratification. In their simulations, Steinmetz \& H{\"o}flich  tracked the evolution of the observed polarization from the electron-scattering photosphere, predicting a relatively high continuum polarization at early times, while the photosphere is in the outer envelope ($P\approx0.4$--1.5\% during the first 10 days), followed by a rotation in $\theta$ and diminishing $P$ as the photosphere passes into the respective prolate and spherically stratified layers of the interior. 

Although the detailed outcome depends on the stellar radius, detailed density profile, rotation (differential versus rigid), and inclination angle, the basic predictions of the above simulations appear to be more or less consistent with the polarized continuum evolution of SN\,2011dh, and thus might provide some general insight. The future construction of similar hydrodynamic models of shock propagation in rotating and/or deformed YSGs could help put the hypothesis of a connection between progenitor deformity and SN\,IIb explosion geometry on a more quantitative footing, leading to more specific predictions for how particular spectropolarimetric observables and their associated timescales can constrain the physical properties of SNe\,IIb and their progenitors.

\subsubsection{Oblique shock breakout}
Recent calculations by Matzner et al. (2013) have demonstrated how nonradial fluid motions generated by aspherical core collapse can dramatically alter the dynamics and emission of the emergent shock. Depending on the degree of asphericity, the shock front can develop an oblique orientation with respect to the stellar surface over a significant fraction of the star. The subsequent outflow from regions crossed and energised by this shock becomes unsteady, potentially stifling the intensity and duration of the breakout flash, limiting ejecta speeds, and casting matter sideways. Collisions of matter outside the star might also occur and significantly influence the observational features of the resulting transient. 

Matzner et al. (2013) discussed SN\,2011dh in the context of oblique shock breakout, and suggested that the resulting asphericity of the highest-velocity ejecta might explain why the early-time luminosity and radio emission indicated a compact progenitor (Arcavi et al. 2011; Soderberg et al. 2012), even though post-explosion images clearly showed that the SN stemmed from an extended YSG (Van Dyk et al. 2013); the reason for the discrepancy lies in the unrealistic assumption of spherical symmetry. Interestingly, oblique shock breakout in SN\,2011dh might also explain why the SN exhibited the strongest continuum polarization during the earliest stages, since the photosphere was in the outer envelope of the quasihomologous flow where velocities are highest. This possibility motivates the construction of hydrodynamic models that will yield information on the aspherical explosion geometries that stem from oblique shock breakout, enabling the further development of radiative transfer and scattering models that will predict the polarization evolution for such a scenario.  

\subsubsection{Asymmetric radioactive heating}
Net continuum polarization from a nonaxisymmetric photospheric configuration could also result from an aspherical distribution of radioactive heating and ionization. For many other CC~SNe, clumps or jet-like configurations of $^{56}$Ni rising rapidly from the core have been invoked to explain their spectropolarimetric properties at various phases (Chugai 1992;  Leonard et al. 2001, 2006; Chugai et al. 2005; Maund et al. 2007a, 2007b; Chornock et al. 2011), and their asymmetric nebular line profiles (Filippenko et al. 1994; Spyromilio 1994; Matheson et al. 2000; Milisavljevic et al. 2010; Chornock et al. 2011; Smith et al. 2015), including the late-time nebular spectrum of SN\,2011dh (Shivvers et al. 2013). 

The hypothesis of aspherical radioactive heating is well motivated; indeed, modern three-dimensional (3D) hydrodynamic models of CC explosions have shown that relatively small convective asymmetries initiated by fast-rising plumes of neutrino-heated matter in the core can, within hours, grow into large-scale asymmetries in the heavy-element distribution of the outflow (Hammer et al. 2010; Wongwathanarat et al. 2010). Models for both red and blue supergiant progenitors with $\sim10\,{\rm M}_{\odot}$ H envelopes produce $^{56}$Ni plumes that mix out to velocities up to 3000--5000\,km\,s$^{-1}$ in the homologous flow (Hammer et al. 2010). More relevantly, 2D hydrodynamic models of SNe\,IIb with H envelopes that are several orders of magnitude less massive indicate that bubbles or plumes of $^{56}$Ni could mix farther out to 6500\,km\,s$^{-1}$ (Iwamoto et al. 1997), near the H/He interface (one might expect the mixing to extend to even larger radii in future 3D models of SNe\,IIb, if the relative velocity predictions from the aforementioned 2D and 3D models of H-rich CC~SNe are mirrored in SNe\,IIb).

\begin{figure}
\includegraphics[width=3.33in]{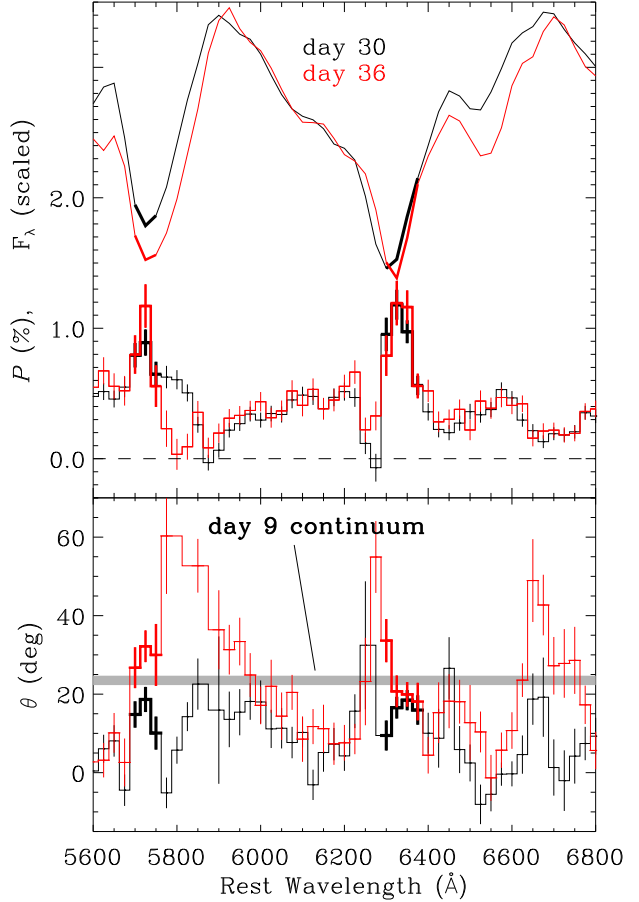}
\caption{Evolution of $P$ and $\theta$ for H$\alpha$ from days 30 and 36. The value of $\theta$ for the day-9 continuum is represented by the horizontal grey line in the lower panel. Spectral regions of interest are highlight with bold lines.}
\label{fig:evol_ha_he}
\end{figure}

Rising $^{56}$Ni plumes can snowplow surrounding material into a shell that can become Rayleigh-Taylor (RT) unstable, potentially mixing ambient envelope material into the bubbles as they rise (Basko 1994). Depending on the path length for the thermalization of gamma rays from the decay of $^{56}$Ni or $^{56}$Co, radioactive plumes will be encased in cocoons of enhanced free-electron density and excess Thomson scattering (e.g., see Chugai et al. 2005). If the cocoons extend into the H layer, they could produce an irregular or mottled photospheric geometry at early times, thus contributing to the net continuum polarization (e.g., on days 9 and 14). Depending the projected position of a given plume relative to the global underlying photosphere, the local polarization integrated over the plumes spheroidal surface could effectively cancel out, having essentially the same effect as a photosphere that is partially obscured by line absorption. This interpretation potentially explains the rough alignment between the position angle of the early day-9 continuum axis and that of H and He\,{\sc i} line polarization on days 30--36, which is further illustrated by Figure\,\ref{fig:evol_ha_he}. In this scenario, we suggest that the same spheroidal plumes that generate locally enhanced Thomson scattering at early times (and where the integrated polarization locally cancels out because of the spherical pattern on the plume) later become plumes of relatively high line optical depth for H$\alpha$ and He\,{\sc i} absorption, since the enhanced electron density in those regions also helps to populate the upper atomic levels of the transitions (e.g., Lucy 1991; Fassia et al. 1998; Chugai 1992, 2006; Chugai et al. 2005; Swartz et al. 1993; H{\"o}flich, Khokhlov, \& Wang 2001). Thus, the plumes at early times will break the symmetry of the global photosphere in a geometric configuration that is similar to symmetry breaking by line absorption at later times, in both cases producing net polarization with similar position angles. Of course, the construction of detailed 3D scattering models that account for the geometries discussed above are needed to put this hypothesis on a more quantitative footing. 

An aspherical distribution of plumes in the ejecta might also explain the knee feature on the red sides of the He\,{\sc i} $\lambda$5876 and H$\alpha$ emission profiles observed on day 30 and later, which could be an imprint of line emission from such structures on the receding hemisphere of the explosion, but just outside the limb of the projected photosphere on the sky. Interestingly, similar morphology was seen in the H$\alpha$ line of SN\,1987A on day 71, interpreted as the imprint of clumpy ejecta on the receding hemisphere of the explosion (Chugai 1992). In that case, moreover, Thomson scattering of photospheric radiation by radioactive plumes on the receding hemisphere of the explosion was successfully modeled as the source of polarization in SN\,1987A.

\begin{figure*}
\includegraphics[width=5.7in]{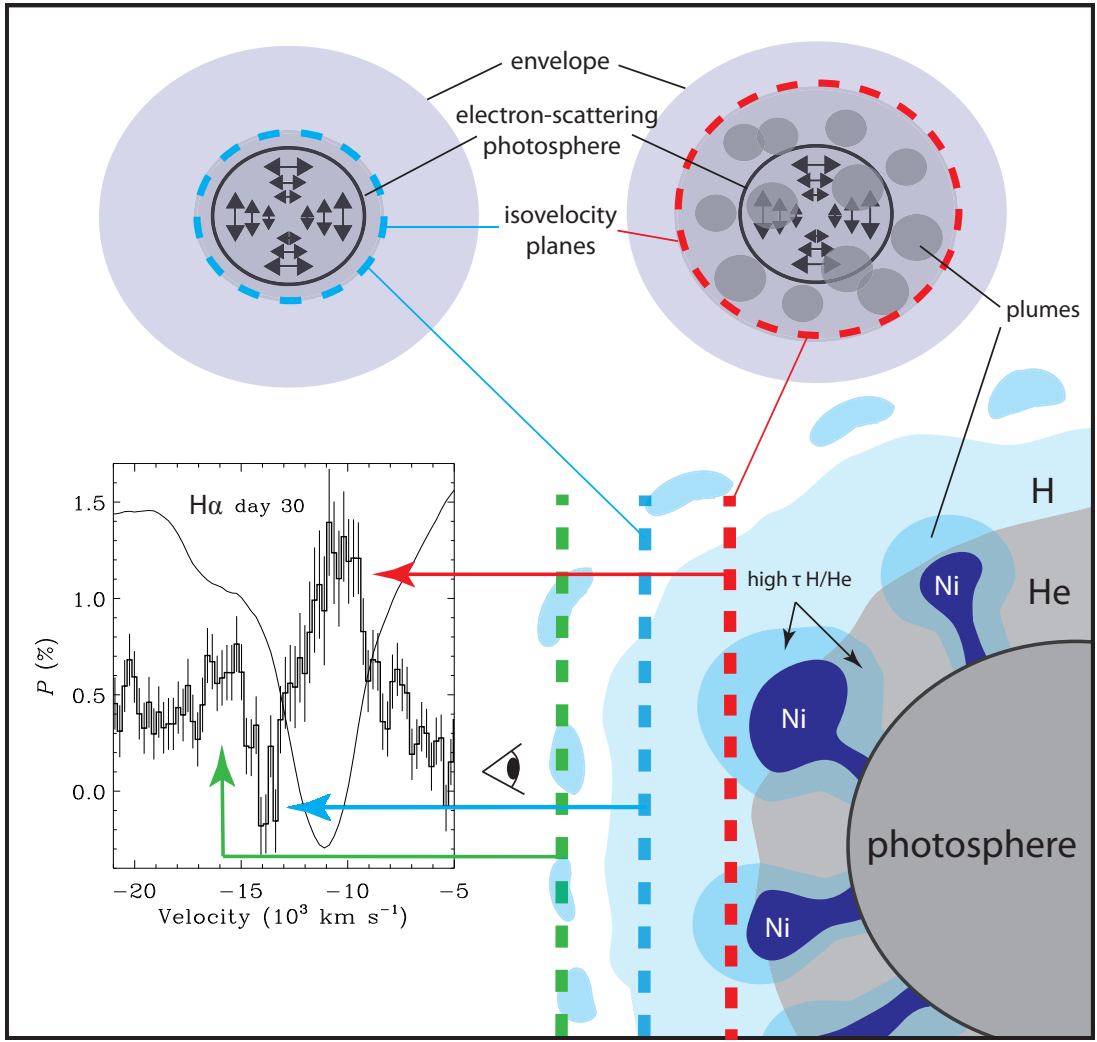}
\caption{Illustration of polarized line formation in SN\,2011dh (not to scale). The formation of the polarized H$\alpha$ profile is represented here (inset at lower left, also showing flux as thinner lines), but the basic idea is applicable to other chemical species that exhibit enhanced line polarization. The upper diagrams are face-on views, and the lower diagram illustrates the transverse view. In the homologous flow, isovelocity planes (dashed lines) are perpendicular to the line of sight. In the face-on views above, isovelocity planes having the largest radial velocity slice through a smaller ``cap" of the sphere closest to the observer, which is why the circular region encompassed by the dashed line is smaller for higher blueshift.   The fastest $-16,000$\,km\,s$^{-1}$ isovelocity plane (green dashed line) intersects the patchy tenuous outer regions of the flow, which asymmetrically obscures the electron-scattering photosphere (arrow-marked region) and give rise to net polarization at those radial velocities. A relatively smooth or tenuous region of the hydrogen envelope is intersected by the $-14,000$\,km\,s$^{-1}$ isovelocity plane (blue dashed line), resulting in uniform or weak obscuration of the photosphere and, hence, low polarization. The $-10,500$\,km\,s$^{-1}$ isovelocity plane (red dashed line) intersects several high optical depth ($\tau$) regions of H$\alpha$/He\,{\sc i}, where ``cocoons" of enhanced excitation for these transitions form around plumes of radioactive $^{56}$Ni, which unevenly obscure the underlying photosphere, giving rise to the peaks in the polarized line profiles.}
\label{fig:cartoon}
\end{figure*}

\begin{figure*}
\includegraphics[width=3.2in]{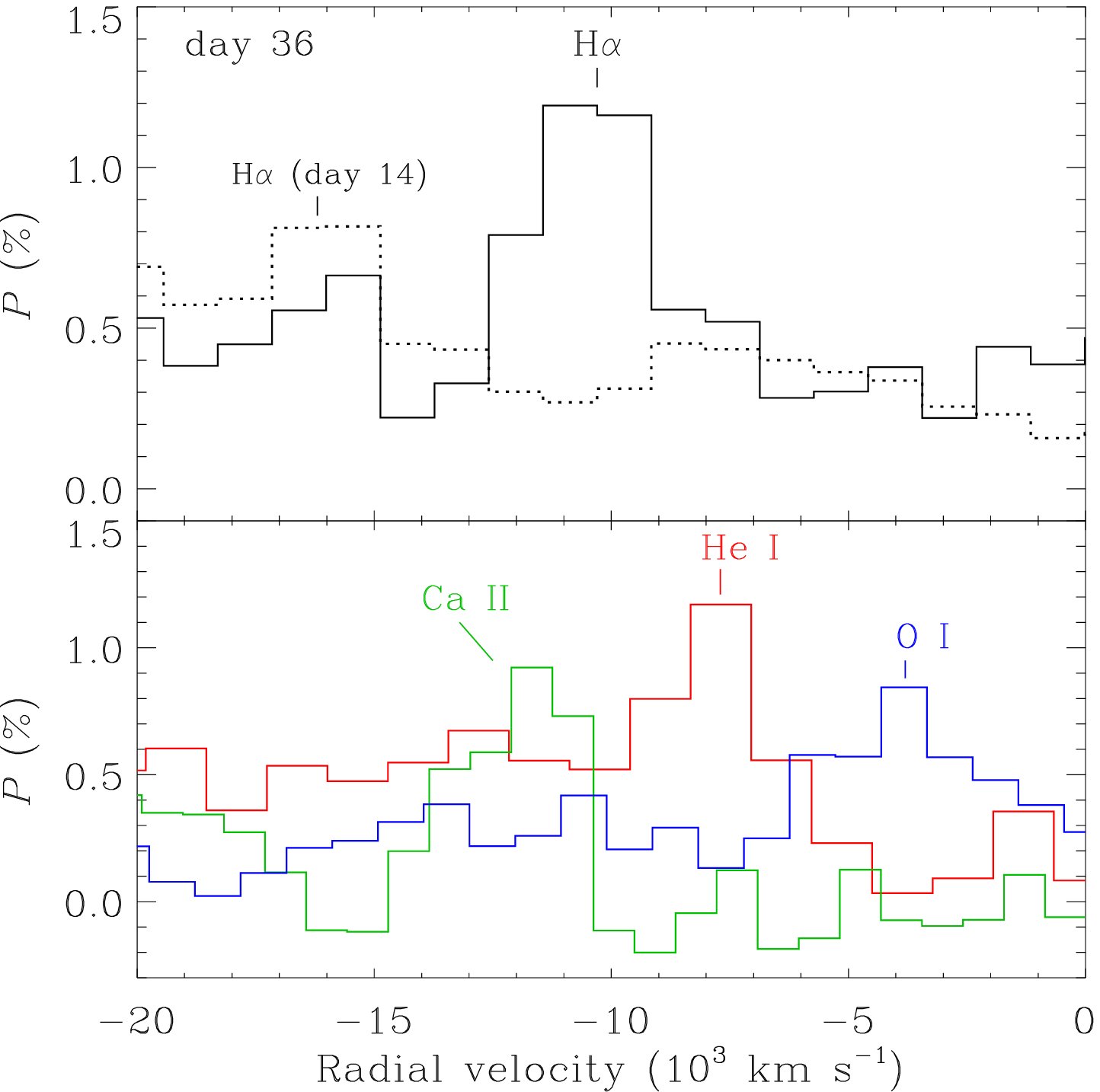}
\includegraphics[width=3.2in]{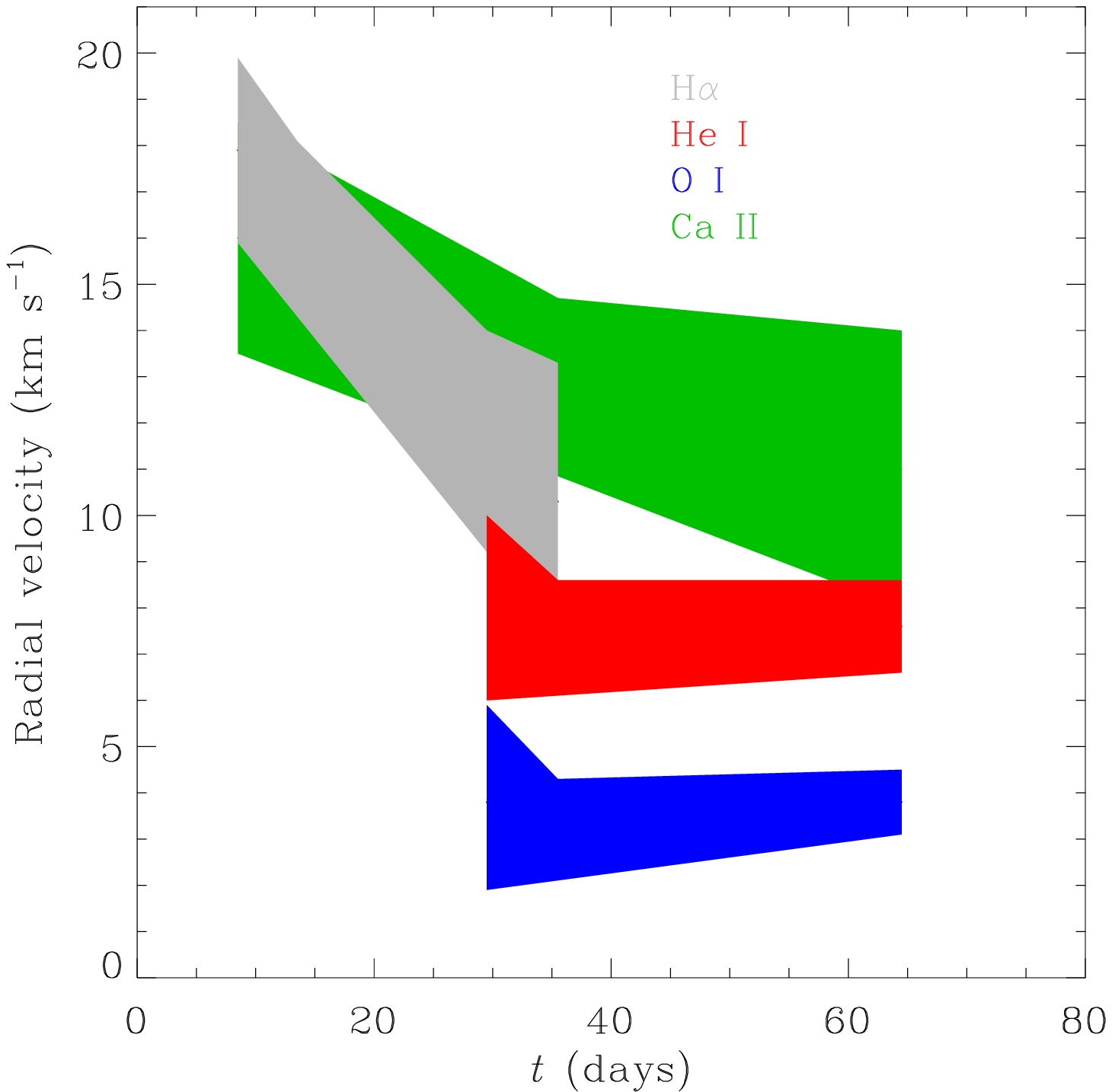}
\caption{\textit{Left}: Enhanced polarization as a function of radial velocity for H$\alpha$, He\,{\sc i}, Ca\,{\sc ii}, and O\,{\sc i} on day 36 (and H$\alpha$ on day 14; dotted line). Homologous expansion is indicated by the relatively fast velocity for H$\alpha$ and Ca\,{\sc ii}, compared to the intermediate velocity for He\,{\sc i} and the slow velocity for O\,{\sc i}. \textit{Right:} Temporal evolution of the radial velocity for the polarized line peaks.}
\label{fig:polvel_rad}
\end{figure*}

Figure\,\ref{fig:cartoon} illustrates a configuration that potentially explains the morphology of the polarized line profiles in the quasihomologous outflow. Homology implies that an imaginary plane intersecting the flow perpendicular to our line of sight (in the plane of the sky) will sample a region of constant RV over that entire plane; we refer to these imaginary cross sections as ``isovelocity planes." The peak RV of line polarization is a useful indication of the depth of patchy absorption by clumps in the quasihomologous flow, and can be used to probe the relative geometries of atomic species at various radii (Kasen et al. 2003). For example, the fastest outermost isovelocity planes with RV $\approx -16,000$\,km\,s$^{-1}$ intersect patches of high-velocity ejecta in the outermost regions of the flow, possibly consisting of clumps from the progenitor star's wind swept up and accelerated by the explosion. These patches partially obscure the polarized photosphere and produce enhanced polarization at those corresponding RV values. Deeper in the homologous flow, isovelocity planes with RV $\approx -14,000$\,km\,s$^{-1}$ intersect a relatively smooth or tenuous region in the thin H envelope that obscures the photosphere evenly or weakly and thus produces no net polarization (creating the steep drop and local minimum of the polarized profile near this value of RV).  Deeper isovelocity planes with RV $\approx -10,500$\,km\,s$^{-1}$ intersect the clumpy region produced by the $^{56}$Ni plumes, creating the peak of the polarized H$\alpha$ and He\,{\sc i} line profiles at this RV.

\subsubsection{Outer Ca\,{\sc ii} and inner O\,{\sc i}}

As illustrated by Figure\,\ref{fig:polvel_rad},  H$\alpha$ and Ca\,{\sc ii} absorption on day 9 exhibit the highest maximum RV observed in SN\,2011dh. At this time, the emission peaks of both lines have relatively high blueshifted velocities of 2000--3000\,km\,s$^{-1}$, with H$\alpha$ moving back toward rest velocity by day 14 (note that the wavelength range of the day-14 data does not sample Ca\,{\sc ii}). Such early blueshifts are a common feature of SNe\,II; the phenomenon has been attributed to the steep density profile of the outflow, which results in highly localised emission regions and exacerbates obscuration of the receding hemisphere of the explosion, resulting in a net blueshift of emission peaks during the first month or so (e.g., see Anderson et al. 2014, and references therein); the relatively low-H envelope masses in SNe\,IIb might explain why the early blueshift lasts for $<14$ days. It thus seems plausible that highest-velocity Ca\,{\sc ii} in the outermost regions of the flow is, like high-velocity H$\alpha$, part of the progenitor's outer envelope or wind swept up and accelerated by the explosion. The apparent antisymmetry of the polarized Ca\,{\sc ii} absorption feature with respect to the day-9 continuum (see Figures\,\ref{fig:rsp} and \ref{fig:ca_polvel}) is consistent with this interpretation, demonstrating that the patchy absorption on day 9 does not share the geometric configuration of the continuum opacity source or the prominent polarized features of H$\alpha$ and He\,{\sc i} on later days. Meanwhile, the disappearance of the high-velocity absorption components and the shift of emission peaks back toward rest velocity indicates that this separate outer region has become transparent by day 14.

O\,{\sc i} exhibits the lowest RV and a significantly different position angle than the other lines, and thus appears to be the hallmark of core oxygen synthesised by the progenitor star during post-main-sequence evolution. Interestingly, the RV of O\,{\sc i} is consistent with the results of modern 3D hydrodynamic simulations of CC explosions, which predict core O to be mixed out to the 4000--5000\,km\,s$^{-1}$ layers of the homologous flow (Hammer, Janka, \& {M{\"u}}ller 2010). Moreover, the different position angle indicates that the inferred clumpiness of O\,{\sc i} could have a different physical origin than that of H$\alpha$ and He\,{\sc i}, perhaps arising from fluid instabilities deeper in the outflow. For SNe\,IIb with low-mass H envelopes, large-scale RT instabilities are unable to develop at the H/He interface, since there is an insufficient overlying mass to facilitate their growth. RT structure is, however, expected to develop at the deeper He/C$+$O interface (Iwamoto et al. 1997), and thus might be another potential source of patchy O\,{\sc i} absorption.  

\subsubsection{The potential effect of scattering by CSM}
Another possibility that should be considered for producing continuum polarization is scattering of SN photons by CSM. If the CSM is sufficiently dense, strong interaction will give rise to an additional vector component of Thomson scattering near the interface of the SN shock and CSM (Hoffman et al. 2008; Mauerhan et al 2014); in the case of dense CSM, however, one might expect the total-flux spectrum to develop narrow emission-line profiles from CSM that becomes photoionized by the UV/X-ray emission at the shock/CSM interface (i.e., this should produce a SN\,IIn spectrum). No such narrow lines were seen in our flux spectra of SN\,2011dh on day 9, or in spectra from others obtained as early as day 4 (Marion et al. 2014), so Thomson scattering from dense ionized CSM seems to be an unlikely alternative for producing an aspherical electron-scattering pseudophotosphere during the first month of our spectropolarimetric data. However, Thomson scattering from CSM interaction cannot be ruled for the earliest phases of SN\,2011dh, where spectropolarimetric data are not available. Indeed, the very early radio and millimeter detection of SN\,2011dh just 3 days post explosion indicates substantial interaction between the SN shock the extended stellar wind of the progenitor (Horesh et al. 2013). Recent ``flash spectroscopy" of other CC~SNe, from just several days to as early as hours after explosion, have indeed exhibited evidence for short-lived CSM interaction and Thomson scattering via Lorentzian emission-line profiles (e.g., Gal-Yam et al. 2014; Shivvers et al. 2015; Smith et al. 2015). Therefore, it is plausible that SN\,2011dh produced a detectable spectropolarimetric signal from CSM interaction before our earliest day-9 data.

Alternatively, small particles of dust in CSM could result in Rayleigh scattering of the optical SN photons (i.e., a ``light echo" effect; Chevalier 1986), which will linearly polarize a substantial fraction of the reflected light. As argued by Wang \& Wheeler (1996) in the case of SN\,1987A, if the dust sublimation timescale is longer than the duration of the UV flash from the SN, small surviving grains can produce an effective medium for optical scattering. This was considered to be a very likely explanation for the strong continuum polarization detected from SN\,1993J (H\"{o}flich et al. 1996; Tran et al. 1997). As mentioned above, early radio and millimeter observations of SN\,2011dh provided direct evidence for CSM from the progenitor wind (Horesh et al. 2013), while late-time X-ray emission roughly 500 days after explosion indicated the presence of dense CSM at $\sim8\times10^{16}$\,cm (Maeda et al. 2014), also consistent with a steady-state YSG wind. The corresponding light travel time across this radius is $\sim31$ days. Therefore, continuum polarization via optical scattering is plausible for the first month of SN\,2011dh if enough dust grains in the wind survived the UV flash of the SN. However, in this case the apparent drop in the continuum polarization near the spectral transition from H-rich to He-rich would have to be a simple coincidence unrelated to the photospheric evolution. 

A potential factor that could help distinguish between Rayleigh scattering by small dust particles and Thomson scattering by free electrons is the wavelength dependence of $P$ for the former case, owing to the wavelength dependence of the Rayleigh scattering phase function. For Thomson scattering, on the other hand, $P$ is wavelength independent. A Rayleigh continuum component, if present, will be most prominent at wavelengths $< 4500$\,{\AA}. The day 14 data from Bok/SPOL provide coverage in the range 4000--4500\,{\AA}, where the average polarization is $P=0.09\% +/- 0.03\%$, nearly null and substantially lower than the continuum level at this epoch. This appears to be inconsistent with Rayleigh scattering by dusty CSM. Instead, the observed characteristics seem consistent with the ``depolarizing" effect of line blanketing in this region of the spectrum by material above the Thomson-scattering photosphere. We therefore conclude that the day 9--14 continuum polarization probably does not include substantial contribution from Rayleigh scattering by dusty CSM.

\subsection{Comparison to other SNe\,IIb} 
Despite their relatively low rates among CC explosions, SNe\,IIb are becoming relatively well represented in terms of objects for which there is high-quality spectropolarimetry. Although thorough temporal sampling is often not obtained, the growing sample is beginning to facilitate a search for systematic spectropolarimetric trends that might point to common physical scenarios for SNe\,IIb and other classes of SNe as well.

\begin{figure}
\includegraphics[width=3.4in]{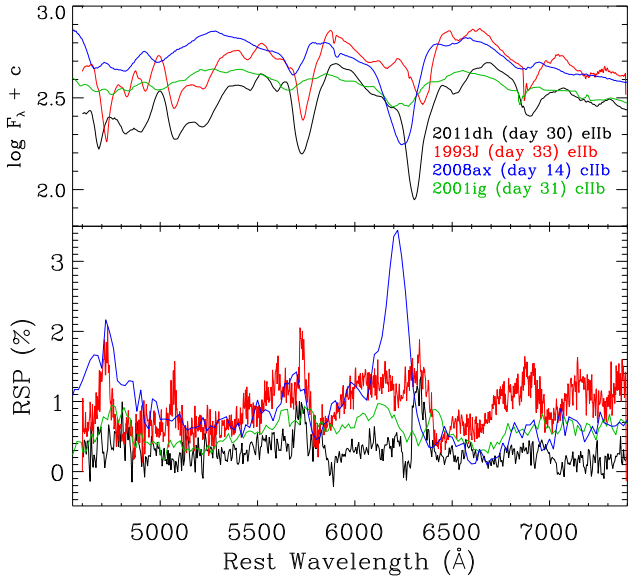}
\caption{Comparison of the total-flux spectra and RSP (rotated by constant angles defined by continua axes) for the eIIb SNe\,2011dh (day 30; black) and 1993J (day 33; red; Tran et al. 1997), and the cIIb SNe\,2008ax (day 14; blue; Chornock et al. 2011) and 2001ig (day 31; green; Maund et al. 2007).}
\label{fig:compare93j}
\end{figure}

Figure\,\ref{fig:compare93j} compares the total-flux spectra and polarization of the eIIb SNe\,2011dh (day 30) and 1993J (day 33; Tran et al. 1997), and the cIIb SNe\,2001ig (day 31; Maund et al. 2007a) and the Ib$\rightarrow$IIb transitional SN\,2008ax (day 14; Chornock et al. 2011), all near phases where He\,{\sc i} has recently become prominent in the spectrum. SN\,2011dh and SN\,1993J exhibit the maxima of their enhanced H$\alpha$ and He\,{\sc i} line polarizations at the same wavelengths and, hence, the same velocities, while SN\,1993J exhibits slightly stronger values of polarization for H$\alpha$ and substantially stronger polarization for He\,{\sc i}. Relatively strong continuum polarization is also evident for SN\,1993J in several regions of the spectrum, while the continuum polarization of SN\,2011dh and SN\,2008ax has diminished before the phases represented in Figure\,\ref{fig:compare93j}, but has yet to rise in the case of SN\,2001ig (see Maund et al. 2007a). SN\,2008ax stands out as having much stronger peak and higher RV H$\alpha$ polarization than the others, by a large margin. 

The conjecture that most SNe\,IIb exhibit substantial continuum polarization at early phases (typically near 0.5\%), relative to the frequent absence of continuum polarization in the early stages of SNe\,II-P, is an interesting trend that could help to reveal important clues about the nature of these explosions and their progenitors. Indeed, if most SNe\,IIb stem from close interacting binary systems, then the imprint of tidal deformation on the continuum polarization (see \S4.1.1) might be present in many cases. Tidal effects were suggested for SN\,1993J to potentially explain its inferred ellipsoidal photosphere (H{\"o}flich et al. 1996; Trammell et al. 1993). Indeed, the progenitor had a radius that was $\sim3$ times larger than that of SN\,2011dh, so the influence of tidal distortion on the spectropolarimetry could potentially be more extreme and longer lasting in that case. For the cIIb subset of SNe that stem from relatively compact progenitors (perhaps, stripped-envelope Wolf-Rayet stars), direct tidal stretching of the progenitor envelope might seem less likely, unless the binary separations are relatively small. If cIIb progenitors are relatively fast rotators, however, then the resulting aspherical density profile for the star could also lead to a similar geometric evolution for the SN shock and the subsequent photosphere (direct tidal deformation by a binary companion, or fast rotation, will both give rise to a steeper density gradient along the polar axis of the star; see \S4.1.1). Interestingly, the transitional cIIb SN\,2008ax exhibited significant continuum polarization on days 6 and 9 ($P\approx0.6$\%), indicating that the outer envelope was indeed quite aspherical; rotational deformation thus might be a plausible explanation in this case. For the separate case of the cIIb SN\,2001ig, however, the evolution from relatively weak continuum polarization during the H-rich phase ($P\approx0.2$\%) to a strong value of $\sim1$\% after the photosphere passed into the He layer (Maund et al. 2007a) is behaviour more in line with observations of SNe\,II-P.  

Alternatively, the effects of aspherical radioactive heating by jets or plumes of $^{56}$Ni might also explain the commonality of continuum polarization in SNe\,IIb. The earlier appearance of this effect, relative to SNe\,II-P with massive H envelopes, is naturally explained by the relatively low-mass envelopes of SNe\,IIb. Furthermore, the apparent connection between the position angles of the early-time continuum, H$\alpha$, and He\,{\sc i} line polarizations in SN\,2011dh, if real, might provide an additional reason to favour this physical scenario (see \S4.1.3). The same connection was proposed for SN\,2008ax, but in that case the similarity between the line and continuum position angles was even more obvious (the system exhibited a higher degree of axisymmetry) and was shared by the O\,{\sc i} line as well (Chornock et al. 2011). The possibility that SN\,2011dh and SN\,2008ax are similar in this regard, even though they are members of the different eIIb and cIIb subclasses, might provide a reason to favour aspherical radioactive heating as the cause of their early-time continuum polarization. After all, the origin of radioactive plumes is in the core, and they will thus be generated regardless of the envelope mass (although the envelope can influence their geometric development). Since line polarization indicates patchy absorption, the comparatively high strength of H$\alpha$ polarization in SN\,2008ax could be related to the compactness and relatively thin H envelope of the progenitor, whereby plumes of $^{56}$Ni from the core might retain their jet-like morphology with less overlying mass to decelerate and mix them, giving rise to the higher degree of axisymmetry. Alternatively, the relatively fast radial velocities of line polarization in SN\,2008ax might also imply that a $^{56}$Ni plume or jet was more highly aligned with our line of sight and thus preferentially obscured the more forward-scattered (less polarized) photons near the apparent centre of the photosphere, giving rise to a higher degree of line polarization. Of course, such viewing-angle effects could potentially blur the evidence for any intrinsic geometric similarities that SNe\,IIb as a class might share, whether they be tidal deformation effects, aspherical plumes of radioactive $^{56}$Ni, or something else (e.g., the potential effect of oblique shock breakout discussed in \S4.1.2).

Comparison of different SNe\,IIb is further complicated by the potential presence of an additional component of continuum polarization from scattering of optical photons from dust in pre-existing CSM. Indeed, SN\,1993J exhibited a relatively large UV excess on day 19, attributable to a substantial amount of CSM, while SN\,2011dh exhibited insignificant UV excess at a similar phase (Ben-Ami et al. 2015). Scattering by dust might thus be the most plausible explanation for the relatively large and long-lasting continuum polarization in SN\,1993J, a possibility that was discussed previously by H\"{o}flich et al. (1996) and Tran et al. (1997). The more recent SN\,2013df also appears to have had a substantial UV excess from CSM (Ben-Ami et al. 2015), and it will be interesting to see whether the spectropolarimetric data of that event share any noteworthy similarities or differences with SN\,1993J and SN\,2011dh that might be attributable to different amounts of CSM in each case. If hypothetically added to SN\,2011dh, another component of continuum polarization from dust scattering could have potentially wiped out the approximate symmetry between the position angles of the continuum polarization at early times and line polarization at later times. Complementary UV data could thus provide important constraints for the interpretation of spectropolarimetry, helping one to anticipate the potential effects of CSM dust.  

\subsubsection{Insight from Cas A}
The scenario of an aspherical SN\,IIb explosion influenced by jets or plumes of enhanced radioactive heating is consistent with the detailed structure of the Cas~A supernova remnant (SNR) and the explosion itself. Spectra of Cas~A light echoes have unambiguously shown that this was a SN\,IIb (Krause et al. 2008), and the substantial differences in the line strength and velocity from multiple echoes that mirrored the SN from different viewing angles is direct evidence for an aspherical explosion (Rest et al. 2011). Interestingly, the SNR also exhibits a highly aspherical distribution of O and Fe knots, which are thought to have been produced by jets or plumes of radioactive $^{56}$Ni that penetrated the outer layers of the SN at early times, or from the growth of RT instabilities (Milisavljevic \& Fesen 2013, 2015). In addition to our spectropolarimetry, the strong multi-peaked [O\,{\sc i}] emission detected from SN\,2011dh at late times by Shivvers et al. (2013) provides independent evidence that an aspherical O-rich SNR like Cas~A could be developing. Chornock et al. (2011) drew similar parallels between Cas~A and SN\,2008ax.

\section{Summary and Concluding Remarks}
We present seven epochs of spectropolarimetry of the Type IIb SN\,2011dh in M51. The observations cover 86 days of evolution, from the early H-rich stage through the nebular phase. This represents the most complete spectropolarimetric  coverage obtained for a SN\,IIb since SN\,1993J. The late-phase polarization at the central wavelengths of nebular features and the $q$--$u$ loops of Ca\,{\sc ii} IR lines at all phases indicate a low value of total ISP ($\lesssim0.1$\%), and we have thus made no attempt to remove it from the data. At each epoch, considerable enhanced polarization is detected across line features, while substantial continuum polarization is firmly detected on days 9 and 14. Several physical scenarios could potentially explain the spectropolarimetric evolution, summarised as follows. 

The global asphericity of the early-time photosphere inferred from the continuum polarization could be the imprint of tidal deformation of the YSG progenitor by the purported massive companion; this hypothesis is facilitated by the recent detection of a hot star at the explosion site by Folatelli et al. (2014), while substantial geometric deformation of the progenitor envelope is consistent with physical parameters of the binary suggested by Benvenuto et al. (2013). This hypothesis does not, however, provide a clear explanation for the approximate symmetry between the early continuum polarization and the subsequent H$\alpha$ and He\,{\sc i} line polarization that we observe  (see \S4.1.3). 

Alternatively, the continuum and line polarization could be the imprint of clumpy excitation of the envelope by fast-rising plumes of $^{56}$Ni from the core. This interpretation does has the advantage of explaining the apparent symmetry between the early-time continuum polarization and the subsequent H$\alpha$ and He\,{\sc i} line polarization. We currently lean toward this scenario (at least for explaining the line polarization), as jets and plumes of $^{56}$Ni are well motivated by hydrodynamic models (Iwamoto et al. 1997; Hammer et al. 2010; Wongwathanarat et al. 2010) and have commonly been invoked to explain the spectropolarimetric and nebular-phase data of other CC~SNe, and in the interpretation of detailed structure in the Cas~A SNR. 

For O\,{\sc i} polarization, the independent geometry and kinematics suggest that the inferred clumpiness of the absorption could be the result of RT fluid instabilities that develop near the He/C$+$O interface, which is consistent with predictions from hydrodynamic models of SNe\,IIb (Iwamoto et al. 1997). For Ca\,{\sc ii} IR polarization, the independent geometry and relatively high-RV edge of the polarized line profile at early times potentially suggests an origin in the outermost regions of the outflow, perhaps in the clumpy swept-up wind of the progenitor.

Another potential explanation for the early-time continuum polarization is optical scattering of SN photons by dusty CSM. Indeed, the presence of extended CSM is indicated by X-ray observations of SN\,2011dh out to day $\sim500$, consistent with ongoing interaction between the reverse shock and CSM produced by a steady-state progenitor wind (Maeda et al. 2014). However, if dust scattering is responsible for the continuum polarization, then it is puzzling why SN\,2011dh has such a meager UV excess compared with SN\,1993J (Ben-Ami et al. 2015), since a mass of CSM capable of scattering the radiation and producing strong continuum polarization should have produced a non-negligible UV excess. Furthermore, the fact that our shortest-wavelength coverage of the polarization on days 14 and 58 does not exhibit an systematic increase in the magnitude of polarization appears to be inconsistent with scattering of SN photons off of dusty CSM. Such scattering by dust could have been mitigated if grains in the CSM were effectively destroyed by the radiation from the SN during the early phases of the explosion.

An aspherical explosion stemming from oblique shock breakout is another possible source of continuum polarization in SN\,2011dh.  
Theoretical calculations indicate that this phenomenon might be relatively common for stripped-envelope SNe (Matzner et al. 2013); perhaps this could explain the commonality of polarized flux from such SNe. 

Although the above interpretations of the spectropolarimetry are plausible and guided by theoretical models, more sophisticated quantitative analysis is required to further discriminate between them and explore potential combinations of their effects. Specifically, the construction of more detailed and robust hydrodynamic and electron-scattering models that account for the geometries we have discussed will enable more accurate interpretations of the emergent polarization.  

SNe\,IIb are not as observationally homogeneous as previously thought. However, some interesting and potentially illuminating patterns appear to be emerging for this class. The frequent detection of substantial polarization demonstrates that asphericity is common, and thus that the assumption of spherical symmetry is dubious; this is probably the reason why SN\,2011dh was classified as a cIIb subtype on the basis of its early-time luminosity and radio emission (Arcavi et al. 2011), even though the explosion clearly stemmed from an extended YSG (Van Dyk et al. 2013). Additional observations of the cIIb and eIIb subclasses undergoing the transition from H-rich to He-rich phases, and complementary UV data indicating the potential for CSM interaction and dust scattering, will be necessary to more firmly establish any systematic trends in their spectropolarimetric properties and to determine the physical reasons for their differences. As it stands, explosion energetics, radioactive inhomogeneities, the mass and potential deformity of stellar envelope, optical scattering by CSM dust, the size and distribution of absorbing clumps relative to the projected photosphere, and viewing-angle effects, are all likely to be important factors in the increasing spectropolarimetric variety of SNe\,IIb. 

\section*{Acknowledgements}
\scriptsize 
We thank the anonymous referee for their insightful commentary and helpful suggestions. We are grateful to the staffs at Lick, Palomar, and Steward Observatories for their excellent assistance, as well as J. Chuck Horst, Julienne Sumandal, and Chris Salvo for help with the Palomar observations. Hien Tran and Ryan Chornock supplied spectropolarimetric data on SN\,1993J and SN\,2008ax for our comparison. J.C.M. acknowledges Dan Kasen and Paul Duffell at U.C. Berkeley for insightful discussions. A.V.F.'s group at U.C. Berkeley is supported by Gary \& Cynthia Bengier, the Richard \& Rhoda Goldman Fund, the Christopher R. Redlich Fund, and the TABASGO Foundation.  Research at Lick Observatory is partially supported by a generous gift from Google. Support was provided by NSF grants AST-1210599 (U.~Arizona), AST-1211916 (U.C.~Berkeley), AST-1009571 and AST-1210311 (SDSU), and AST-1210372 (U.~Denver). J.M.S. is supported by an NSF Astronomy and Astrophysics Postdoctoral Fellowship under award AST-1302771. A.G.Y. is supported by the EU/FP7 via ERC grant no. 307260, the Quantum Universe I-Core program by the Israeli Committee for Planning and Budgeting and the ISF; by Minerva and ISF grants; by the Weizmann-UK ``making connections'' program; and by Kimmel and ARCHES awards.

\end{document}